\documentclass{aa}  

\usepackage{graphicx}
\usepackage{txfonts}
\usepackage{xcolor}

\begin{document}

   \title{The 21cm--galaxy cross-correlation: Realistic forecast for 21cm signal detection and reionisation constraints}
   
   \author{Anne~Hutter\inst{1,2,3}\fnmsep\thanks{\email{anne.hutter@univie.ac.at}} and Caroline Heneka\inst{4}
          }
    \authorrunning{Hutter et al.}

   \institute{Department of Astrophysics, University of Vienna, T\"urkenschanzstrasse 17, A-1180 Vienna, Austria
         \and Niels Bohr Institute, University of Copenhagen, Jagtvej 128, DK-2200, Copenhagen N, Denmark
         \and Cosmic Dawn Center (DAWN)
         \and Institut f\"ur Theoretische Physik, Universit\"at Heidelberg, Philosophenweg 16, 69120 Heidelberg, Germany 
         }

   \date{Received - -, -; accepted - -, -}

  \abstract
    {The 21cm--galaxy cross-correlations are expected to play a key role in the confirmation of the cosmological 21cm signal. 
    }
    {We investigate which survey configurations detect the 21cm--Lyman-$\alpha$ emitter (LAE) cross-correlation signal and assess its ability to distinguish reionisation scenarios. 
    } 
    {Our pipeline computes observational uncertainties for the 21cm--galaxy cross-power spectrum, accounting for key survey parameters. These include field of view (FoV); limiting luminosity of galaxy surveys, $L_\alpha$; redshift uncertainty, $\sigma_\mathrm{z}$; and 21cm foreground wedge assumptions. Using this pipeline, we calculated the signal-to-noise ratio (S/N) of the 21cm--LAE cross-power spectrum for two scenarios: one where faint galaxies dominate the ionising photon budget and one where reionisation is driven by bright galaxies. 
    }
    {We find that: (i) S/N increases with a larger FoV, fainter $L_\alpha$, and smaller $\sigma_\mathrm{z}$, with the FoV having the strongest impact when $\sigma_\mathrm{z}$ is small. 
    (ii) Under a moderate foreground wedge, photometric-like surveys yield an insufficient S/N, and medium-deep ($L_\alpha\gtrsim10^{42.5}$erg~s$^{-1}$), wide-area ($\mathrm{FoV}>20$~deg$^2$) slitless spectroscopic or spectroscopic surveys are required. 
    (iii) Under an optimistic foreground wedge, detection is possible with deep ($L_\alpha\gtrsim10^{42.3}$erg~s$^{-1}$), wide-area ($\mathrm{FoV}\gtrsim80$~deg$^2$) photometric-like surveys, or with shallower, small-area ($\mathrm{FoV}\simeq2-3$~deg$^2$) slitless spectroscopic surveys.
    (iv) To distinguish the two reionisation scenarios at $z=7$, moderate foreground wedge scenarios require deep-wide spectroscopic surveys; under optimistic foreground wedge, assumptions shallower ($L_\alpha\simeq10^{42.8}$erg~s$^{-1}$), medium-area ($\mathrm{FoV}\simeq 10$~deg$^2$) slitless spectroscopic surveys suffice. 
    (v) Maximising the S/N for both detection and model discrimination requires sampling the large-scale peak of the cross-power spectrum, which shifts to larger physical scales as reionisation proceeds, and the less ionisation fronts follow the gas density, thus making surveys at $z>7$, or when the Universe was more neutral ($\langle\chi_\mathrm{HII}\rangle > 0.5$), more promising despite lower galaxy number densities.}
    {Our results show that large-area spectroscopic surveys and 21cm foreground cleaning are key for using 21cm–LAE cross-correlations to constrain reionisation beyond the global ionisation state of the intergalactic medium.}
   

   \maketitle

\section{Introduction}
\label{sec_introduction}

The first galaxies emerging during the first billion years ushered in the epoch of reionisation (EoR) by emitting ultraviolet (UV) radiation that ionised the surrounding neutral hydrogen gas in the intergalactic medium \citep[IGM;][]{Barkana2001, Dayal2018}. Ionised regions grew around galaxies until the IGM was fully ionised by $z\simeq5.3$ \citep{bosman2022}. However, with the nature of these early galaxies remaining uncertain and JWST providing intriguing but puzzling data \citep[e.g.][]{Labbe2023, Adams2023, Adams2024, ArrabalHaro2023, Atek2023, Austin2023, BoylanKolchin2023, Donnan2024, Harikane2025, McLeod2024, carniani2024, witstok2025, eugenio2025, Cameron2024, Isobe2023, Curti2025, Senchyna2024, Topping2025, Watanabe2024, Nakane2024}, key questions remain regarding how quickly did galaxies grow and become luminous and whether reionisation was driven by the many faint or few bright galaxies.
Answering the latter using galaxy observations alone is challenging, as a direct measurement of ionising radiation from the first galaxies are blocked by the partially neutral IGM during the EoR. Instead, estimates rely on on indirect proxies, such as emission lines and other galaxy properties \citep[e.g.][]{chisholm2020, Jaskot2024a, Jaskot2024b, leclercq2024, Mascia2024}.

A complementary and powerful probe of reionisation is the 21cm signal produced by the hyperfine transition of neutral hydrogen. Radio interferometers such as 
the Square Kilometre Array\footnote{Square Kilometre Array, \url{https://www.skatetelescope.com}} \citep[SKA;][]{Carilli2004}, Hydrogen Epoch of Reionisation Array \citep[HERA;][]{DeBoer2017, HERA2023, Berkhout2024, HERA2025}, Murchison Widefield Array\footnote{Murchison Widefield Array, \url{http://www.mwatelescope.org}} \citep[MWA;][]{Li2019, Barry2019, Trott2020, Trott2025, Nunhokee2025}, and Low Frequency Array\footnote{Low Frequency Array, \url{http://www.lofar.org}} \citep[LOFAR;][]{Patil2017, Mertens2020, Mertens2025} will provide statistical detection and tomographic maps of this signal, enabling us to trace the evolving morphology of ionised regions in the IGM. These maps will offer a way to infer the ionising outputs from galaxies, including those too faint to observe. 
However, detecting the 21cm signal alone is challenging due to foregrounds that are spectrally smooth but several orders of magnitude brighter. These intrinsic foregrounds, primarily originating from our Galaxy and extragalactic point sources, are further complicated by the chromaticity of radio interferometers, which causes them to leak into higher Fourier modes along the line of sight, forming the so-called foreground wedge \citep{Parsons2012b}. Importantly, because these foregrounds are unrelated to the large-scale distribution of galaxies during the EoR, their signals should not correlate \citep{Furlanetto2007, Liu2020}. This makes cross-correlations between 21cm observations and galaxy surveys a promising technique for detecting the cosmological 21cm signal from the EoR \citep{Furlanetto2007, Beane2019, McBridge2024}.

Various studies have investigated the detectability of the 21cm--galaxy cross-correlation signal using combinations of different instruments. These include 21cm experiments such as SKA, MWA, LOFAR, and HERA and galaxy surveys using, for example, Subaru's Hyper Suprime Cam, its Prime Focus Spectrograph (PFS), or the Roman Space Telescope \citep[Roman;][]{Park2014, Sobacchi2016, Hutter2017, Hutter2018b, Vrbanec2016, Vrbanec2020, Kubota2018, Kubota2020, Heneka2017, Heneka2020, Weinberger2020, Yoshiura2018, LaPlante2023, Moriwaki2024, Gagnon-Hartman2025}. Related efforts have also pursued stacking analyses of 21cm maps on samples of LAEs, ranging from simulation-based feasibility studies \citep{Hutter2017, Davies2021, Chen2025} to applications on existing observations \citep{Trott2021}, which constitute an alternative form of 21cm–galaxy cross-correlation.
While most of these studies focused on assessing 21cm--galaxy cross-correlation detectability for specific combinations of instruments and survey designs, a subset has gone beyond fixed survey setups to explore how the signal-to-noise ratio (S/N) of the 21cm--galaxy cross-power-spectrum changes with survey characteristics, such as the survey volume, the galaxy number density, and galaxy redshift precision. However, these analyses typically vary only a subset of these parameters \citep{Hutter2017, Heneka2021, Weinberger2020, Moriwaki2024}, and only a few consider them in combination \citep{Gagnon-Hartman2025}.
A key challenge in survey design are the competing observational requirements: While large-volume surveys help reduce uncertainties in 21cm observations, small, deep surveys better mitigate shot noise and redshift uncertainty inherent in galaxy surveys. To minimise redshift uncertainty, most studies focus on line emitters -- galaxies exhibiting detectable Lyman-$\alpha$ or OIII emission during the EoR.

This tension between survey requirements for 21cm and galaxy observations has been explored in part by previous works: \citet{Hutter2018b} examined how survey volume and minimum Ly$\alpha$ luminosity affect the ability to constrain the global IGM ionisation fraction through 21cm--Ly$\alpha$-emitter (LAE) cross-correlations, while \citet{Heneka2021} showed how the S/N varies with survey depth (sensitivity limit) and spectral resolution for the 21cm--Ly$\alpha$ cross-signal. \citet{Weinberger2020} analysed the S/N of the 21cm--LAE cross-power spectrum as a function of survey volume and minimum detectable luminosity, while \citet{Moriwaki2024} conducted a comparable study on the 21cm--OIII cross-power spectrum, relating S/N to survey volume and minimum halo mass. \citet{Gagnon-Hartman2025} recently included galaxy redshift uncertainty as an explicit variable, highlighting its critical role in survey design. However, none of these studies has systematically explored how the 21cm--LAE cross-power spectrum behaves across a broader range of galaxy luminosities and 21cm foreground wedge models, nor which survey characteristics would be needed to distinguish reionisation scenarios with different ionisation morphologies.

In \citet{Hutter2023b}, we showed that characteristic features of the 21cm--LAE cross-correlation function -- such as its inversion point and small-scale amplitude -- depend on the ionisation morphology by tracing the typical size of the ionised regions around LAEs and the average neutral hydrogen gas density.  We found that the small-scale anti-correlation weakens when ionisation fronts more closely follow the underlying cosmic web density structure. These features are also imprinted in the 21cm--LAE cross-power spectrum: The position of the large-scale negative peak and the scale at which the signal changes sign encode information about the average neutral fraction and the characteristic sizes of ionised regions.

In this paper, we build on these results and investigate the survey characteristics for (a) detecting the 21cm--LAE cross-power spectrum across two orders of magnitude in Lyman-$\alpha$ (Ly$\alpha$) luminosity and for two foreground wedge models, and (b) distinguishing between two physically motivated reionisation scenarios: one in which faint galaxies dominate the ionising photon budget and one where reionisation is driven by bright galaxies. To achieve this, we used the {\sc mhdec} and {\sc mhinc} {\sc astraeus} simulations, which model galaxy evolution and reionisation self-consistently. We varied the survey field of view (FoV), luminosity thresholds, redshift uncertainty, and foreground wedge models to quantify their effect on both signal detectability and sensitivity to ionisation morphology. Our results provide practical guidance for designing future surveys to extract the maximum insight from 21cm--galaxy cross-correlation measurements.

This paper is organised as follows. In Section~\ref{sec_signal}, we discuss the dependence of the 21cm--galaxy cross-power spectrum on reionisation and the selected galaxy sample before we explore how its observational uncertainties depend on the survey parameters and would need to be optimised for detection in Section~\ref{sec_error} and \ref{sec_SNR}. Section~\ref{sec_reionisation_scenarios} explores which surveys would allow us to distinguish between reionisation scenarios demarcating two extremes. We conclude in Section~\ref{sec_conclusions}.

\section{The 21cm--galaxy cross-power spectrum}
\label{sec_signal}

\begin{figure*}
\resizebox{\hsize}{!}
  {\includegraphics{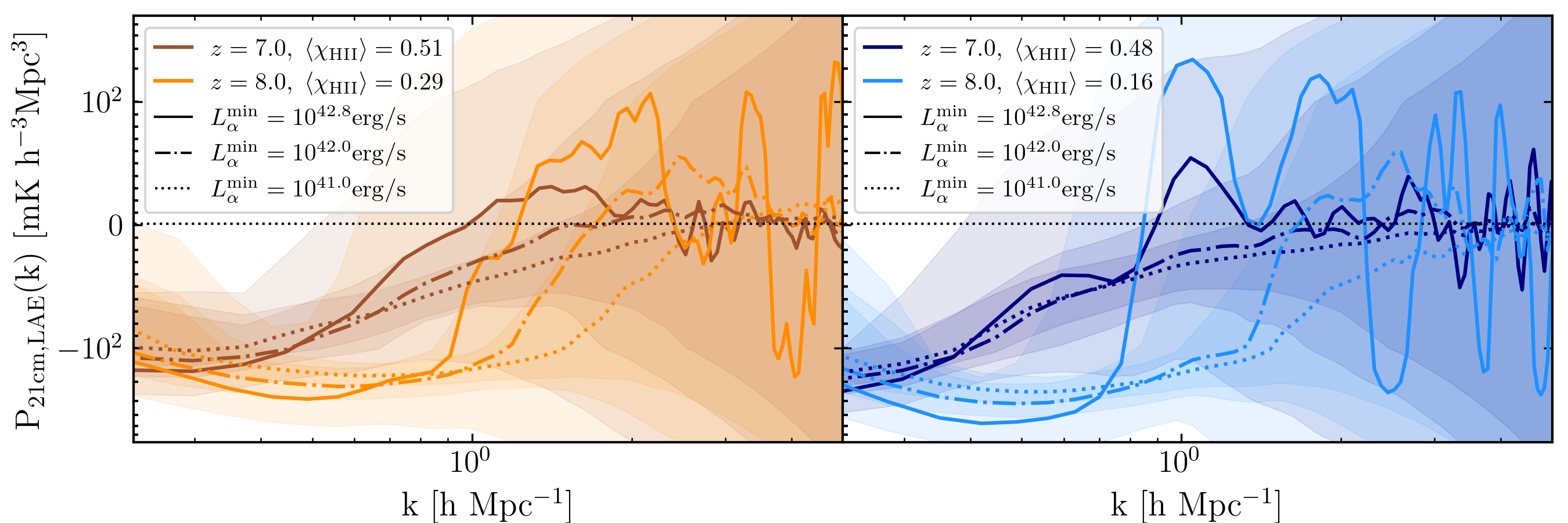}}
  \caption{21cm-LAE cross-power spectrum at $z=7$ and $8$ for the faint galaxy ({\sc mhdec}, left) and bright galaxy ({\sc mhinc}, right) reionisation scenarios~\citep{Hutter2023b} and different minimum observed Lyman-$\alpha$ luminosities of $L_\alpha^\mathrm{min}=10^{41.0}$erg/s (dotted lines), $10^{42.0}$erg/s (dash-dotted lines), $10^{42.8}$erg/s (solid lines). Shaded regions depict the observational uncertainties for a spectroscopic survey ($\sigma_z=0.001$) covering a FoV of $25$~deg$^2$ assuming the SKA-Low1 AA4 antenna layout and a moderate foreground model, see Section~\ref{sec_error}. The cross-power spectrum values shown are interpolated from $65$ and $68$ linearly spaced $k$ bins (following the $k$ binning described in Section~\ref{sec_error}) at redshifts $z=7$ and $8$, respectively, and smoothed with a Gaussian kernel of standard deviation $\sigma=1$~bin.}
     \label{fig_crossps_Lmin}
\end{figure*}

\subsection{Theoretical background of the signal}
\label{sec:PS-signal}

The 21cm--galaxy cross-power spectrum, which traces the 21cm signal distribution around galaxies, is sensitive to the morphology of the 21cm signal maps and characteristics of the selected galaxy population. For this reason, we considered two reionisation scenarios: one driven by faint low-mass galaxies and one by bright massive galaxies, with different detection limits on the Lyman-$\alpha$ (Ly$\alpha$) luminosities of galaxies. The two scenarios investigated in this study correspond to the {\sc mhdec} (faint galaxies) and {\sc mhinc} (bright galaxies) models from \citet{Hutter2023a}, simulated with the {\sc astraeus} framework \citep{Hutter2021a,Hutter2021b} and previously analysed for their 21cm-Ly$\alpha$ emitter (LAE) cross-correlation signal in \citet{Hutter2023b}. 
While \citet{Hutter2023b} focused on differences in the 21cm-LAE cross-correlation functions for the {\sc mhdec} and {\sc mhinc} scenarios, here we extend that work by analysing the corresponding 21cm-LAE cross-power spectra and their dependence on characteristics that connect to observations, such as the assumed limiting Ly$\alpha$ luminosity of detectable galaxies, and, in the following sections, the inclusion of noise modelling for realistic forecasts on signal detection and signal-to-noise.

We computed the 21cm differential brightness temperature $\delta T_b(\vec{x})$ from the simulated density and ionisation fields following \citet{Furlanetto2006},
\begin{eqnarray}
    \delta T_b(\bf{x}) &\simeq& \frac{3c\lambda_{21}^2 h A_{10}}{32 \pi k_\mathrm{B} H_0}\ n_{H,0} \left( \frac{1+z}{\Omega_m} \right)^{1/2} \left( 1-\frac{T_\mathrm{CMB}}{T_\mathrm{s}(\bf{x})} \right) \chi_\mathrm{HI}(\bf{x}) \nonumber \\
    &=& T_0\  \left( 1-\frac{T_\mathrm{CMB}}{T_\mathrm{s}({\bf x})} \right) \chi_\mathrm{HI}({\bf x}) \left( 1 + \delta({\bf x}) \right),
\label{eq_deltaTb}
\end{eqnarray}
with $A_{10}$ the Einstein coefficient for spontaneous emission of a photon with an energy of $hc/\lambda_{21}$, $n_\mathrm{H,0}$ the neutral hydrogen density today, $\chi_\mathrm{HI}({\bf x})$ and $1+\delta({\bf x})=\rho({\bf x})/\langle \rho\rangle$ describing the neutral hydrogen fraction and overdensity at position ${\bf x}$, respectively, and the CMB temperature $T_\mathrm{CMB}$. In the following, we assume the post-heating regime where the spin temperature $T_\mathrm{S}$ is fully coupled to the kinetic gas temperature ($T_\mathrm{s} \gg T_\mathrm{CMB}$) - a reasonable assumption during the mid and end stages of reionisation and valid for $z\lesssim8$ based on recent HERA constraints \citep{HERA2022,HERA2023}. 
We defined the fractional 21cm fluctuation field as
\begin{eqnarray}
    \delta_{21}(\vec{x})&=&\frac{\delta T_b(\vec{x}) - \langle \delta T_b \rangle}{\langle \delta T_b \rangle},
\end{eqnarray}
and constructed a similar galaxy overdensity field from the simulated galaxy distribution as
\begin{eqnarray}
    \delta_{\mathrm{gal}}(\vec{x}) &=& \frac{n_\mathrm{gal}(\vec{x}) - \langle n_\mathrm{gal} \rangle}{\langle n_\mathrm{gal} \rangle}.
\end{eqnarray}
The 21cm--galaxy cross-power spectrum is then given by the product of these two fields in Fourier space,
\begin{eqnarray}
    \langle \delta_{21}(\vec{k}) \, \delta^*_{\mathrm{gal}}(\vec{k'}) \rangle &=& (2\pi)^3 P_{\mathrm{21,gal}}(k) \, \delta_D(\vec{k} - \vec{k'}),
\end{eqnarray}
where $\delta_D$ is the Dirac delta function.\footnote{Note that we assume the following convention for the Fourier transformation: $\tilde\delta(\vec{k}) = \int \mathrm{d}^3x ~ \delta(\vec{x}) ~e^{i\vec{k}\vec{x}}$ and $\delta(\vec{x}) = \frac{1}{(2\pi)^3} \int \mathrm{d}^3k~ \tilde\delta(\vec{k}) ~e^{-i \vec{k} \vec{x}}$} 
To obtain the 21cm--LAE cross-power spectra from the {\sc astraeus} simulations, we (1) derive the Ly$\alpha$ luminosities\footnote{The derived Ly$\alpha$ luminosities reproduce the observed Ly$\alpha$ luminosity functions at $z\simeq7$.} of all galaxies accounting for their simulated star formation histories, the assumed ionising escape fraction $f_\mathrm{esc}$ and the attenuation by dust and neutral hydrogen distribution in the IGM \citep[see Gaussian model in][]{Hutter2023a}; (2) select all galaxies with Ly$\alpha$ luminosities above a given threshold, $L_\alpha^\mathrm{min}$; (3) generate the corresponding overdensity field and map it onto a 3D grid; and (4) compute the 21cm--galaxy cross-power spectra using this 3D grid along with the 21cm fractional fluctuation field derived from the simulated ionisation and density fields.

\subsection{Dependence on reionisation and galaxy sample}
\label{subsec_eor_galaxy}

To illustrate how the 21cm--LAE cross-power spectrum traces the ionisation morphology and history, and how it depends on the underlying galaxy sample, we show in Fig.~\ref{fig_crossps_Lmin} the cross-power spectra for galaxy populations with different Ly$\alpha$ luminosity thresholds, $L_\alpha^\mathrm{min}=10^{41}$, $10^{42}$, $10^{42.8}$~erg~s$^{-1}$, and for two contrasting reionisation scenarios ({\sc mhdec} and {\sc mhinc}) at $z=7$ and $8$ (and thus different ionisation fractions).

\paragraph{Dependence on reionisation:} 
The key features of the 21cm--LAE cross-power spectrum, such as its sign, amplitude and zero-crossing, encode information about the heating and ionisation morphology of the IGM. 

Firstly, at large scales ($k\lesssim 1 h$~Mpc$^{-1}$) and during reionisation, the 21cm and LAE number density fields are anti-correlated, because LAEs reside in ionised regions where the 21cm signal is suppressed. By tracing the average neutral hydrogen density, the amplitude of this large-scale anti-correlation reflects the global ionisation state of the IGM and how ionisation fronts propagate through the cosmic web. When the ionisation field more closely follows the underlying gas density -- as is the {\sc mhdec} scenario, where low-mass galaxies drive reionisation -- this anti-correlation is weaker \citep[c.f. the comparison to the appropriate analytical description in][]{Hutter2023b}. For example, at $z=7$ when both scenarios have a global ionisation fraction of $\langle \chi_\mathrm{HII} \rangle\sim0.5$, the maximum negative amplitude at $z=7$ reaches nearly $250$~mK~$h^{-3}$Mpc$^{3}$ for the brightest galaxy sample ($L_\alpha^\mathrm{min} \geq 10^{42.8}$erg~s$^{-1}$) in the {\sc mhinc} scenario, while it only goes up to $\sim150$~mK~$h^{-3}$Mpc$^{3}$ in the {\sc mhdec} scenario. At higher redshifts, for example at $z=8$, the differences in the maximum negative amplitude are more dominated by different values of the global ionisation fraction besides the correction through gas density (ionisation morphology). 

Secondly, the first zero-crossing scale of the cross-power spectrum provides another diagnostic: it approximately corresponds to the typical size of ionised regions around the selected galaxies \citep{Vrbanec2016, Vrbanec2020, Hutter2017, Heneka2017, Kubota2018, Heneka2020, Hutter2023a}: the larger these regions, the more the anti-correlation shifts to larger scales (smaller $k$). As the IGM becomes increasingly ionised from $z=8$ to $z=7$, the ionised regions grow, shifting the first zero crossing to larger physical scales (smaller $k$ values) in both reionisation scenarios and for all Ly$\alpha$ luminosity thresholds.

\paragraph{Dependence on galaxy sample:}
Since each galaxy survey has its own flux limit that sets the faintest detectable objects, it is important to consider how the 21cm--LAE cross-power spectrum changes when varying this flux limit -- or in our case the limiting Ly$\alpha$ luminosity $L_\alpha^\mathrm{min}$. 
Including fainter (low-mass) galaxies by lowering $L_\alpha^\mathrm{min}$ means the 21cm--LAE cross-power spectrum probes less massive galaxies, which reside in less biased, less overdense, and less ionised environments. This trend is evident in Fig.~\ref{fig_crossps_Lmin}: as $L_\alpha^\mathrm{min}$ decreases from $10^{42.8}$erg~s$^{-1}$ to $10^{41}$erg~s$^{-1}$, the first zero crossing of the 21cm-LAE cross-power spectrum shifts to smaller physical scales (larger $k$ values). Moreover, the difference in the first zero crossing scales between bright and faint samples becomes more pronounced at higher redshifts, since galaxies of a given luminosity trace a more biased population.

\section{The observational uncertainties}
\label{sec_error}

The uncertainty in the 21cm-galaxy cross-power spectrum arises from the finite sampling of Fourier modes (cosmic variance), instrumental noise (e.g. thermal) and survey noise due to survey window functions (e.g. limiting spectral and spatial resolution), shot noise due to a limited galaxy number density, as well as foregrounds. Following \citet{Furlanetto2007, McQuinn2007, Lidz2009}, the total variance of the cross-power spectrum at a given Fourier mode $\vec{k}$ is given by
\begin{eqnarray}
    \sigma^2_\mathrm{21,gal}(\vec{k}) &=& \frac{1}{2} \left[ P^2_\mathrm{21,gal}(\vec{k}) + \sigma_{21}(\vec{k})~\sigma_\mathrm{gal}(\vec{k}) \right], 
\label{eq:sigma_cross}
\end{eqnarray}
where $P_\mathrm{21,gal}$ is the true cross-power spectrum, $\sigma_{21}$ is the total 21cm power spectrum uncertainty (including cosmic variance and thermal noise), and $\sigma_{\mathrm{gal}}$ is the uncertainty on the galaxy power spectrum (cosmic variance plus shot noise). 
The factor of $1/2$ arises from assuming the 21cm and galaxy fields are Gaussian random fields, as is commonly done in analytic estimates of the cross-power spectrum variance. 
We note that Eq.~\ref{eq:sigma_cross} assumes that the 21cm-galaxy cross-power spectrum covariance matrix is purely diagonal. Tests with multiple realisations of the 21cm noise and galaxy fields indicate the assumption is approximately valid (App.~\ref{sec_covariances}). While galaxy shot noise and interferometer thermal noise are uncorrelated, additional mode-coupling effects (e.g. mutual coupling, residual foregrounds) could in principle introduce off-diagonal covariance, which are neglected here.

\paragraph{The 21cm signal error $\sigma_{21}(\vec{k})$:}
The total 21cm power spectrum uncertainty consists of the underlying cosmological 21cm power spectrum, $P_{21}(\vec{k})$, (cosmic variance) and the thermal noise arising from the radio interferometer, $P_\mathrm{noise}(\vec{k})$;
\begin{eqnarray}
    \sigma_{\mathrm{21}}(\vec{k}) &=& P_{21}(\vec{k}) + P_\mathrm{noise}(\vec{k}).
\end{eqnarray}
Here, the thermal noise depends on the array configuration of the interferometer, determined through corresponding baseline density and thus u,v-coverage the number of Fourier modes measured, as well as solid angle $\Omega'$ of the primary beam FoV, system temperature $T_\mathrm{sys}$, and observation time $t$. Following \citet{Pober2014} the thermal noise term is given by
\begin{eqnarray}
    P_\mathrm{noise}(\vec{k}) &=&X^2 Y \frac{\Omega'}{2~t(\vec{k})} T_\mathrm{sys}^2.
\end{eqnarray}
Here $X$ and $Y$ are the comoving distance factors converting angles and frequencies to transverse and line-of-sight distances, respectively (see \citealt{Parsons2012a}). $\Omega' = \Omega_\mathrm{p}^2 / \Omega_\mathrm{pp}$ is the power primary beam solid angle $\Omega_\mathrm{p}$ squared divided by the solid angle of the squared power primary beam $\Omega_\mathrm{pp}$ (see \citealt{Parsons2014} for a derivation),\footnote{When converting from flux density-based visibilities into brightness temperature in mK power spectra, $\Omega_p$ enters into the unit conversion, while $\Omega_{pp}$ arises from the inherent correlation of modes from the power spectrum's definition.} while $t(\vec{k})$ is the total integration time for a given mode $\vec{k}$ which depends on the baseline density distribution of the interferometer array. $T_\mathrm{sys} = T_\mathrm{rcv} + T_\mathrm{sky}$ is dominated at low frequencies by the galactic synchrotron foreground $T_\mathrm{sky} \approx 6\times10^4~ \mathrm{mK} \left( \frac{\nu}{300~\mathrm{MHz}}\right)^{-2.55}$ and by the receiver temperature $T_\mathrm{rcv}= 0.1\, T_\mathrm{sky} + 40~\mathrm{mK}$ of SKA-Low1 otherwise.
Throughout this paper, we assume the SKA-Low1 AA4 antenna layout and include baselines up to $3.4$~km, a total integration time of $1000$~h, a survey bandwidth of $B=8$~MHz, and the moderate and optimistic foreground models as defined in \citet{Parsons2012a}.  
The survey volume is set by the bandwidth, $B$, and the solid angle of its FoV, $\Omega_\mathrm{survey}$, as
\begin{equation}
    V_\mathrm{survey} = X^2 Y \, \Omega_\mathrm{survey} \, B \,.
\end{equation}
At $z=7$, the survey extends along the line of sight by $\sim125$~cMpc, corresponding to $\Delta z \simeq 0.36$. 
While our results use the AA4 layout, we have also computed all forecasts with the AA$\star$ layout for SKA-Low1 science verification, finding only minimal reductions in sensitivity compared to AA4.

\paragraph{The galaxy survey error $\sigma_\mathrm{gal}(\vec{k})$:}
The galaxy power spectrum uncertainty can be expressed as
\begin{eqnarray}
    \sigma_{\mathrm{gal}}(\vec{k}) &=& P_{\mathrm{gal}}(\vec{k}) + P_\mathrm{shot}(\vec{k}) , 
\end{eqnarray}
where $P_\mathrm{gal}(\vec{k})$ is the cosmological galaxy power spectrum and reflects cosmic variance, i.e. fluctuations due to the finite number of Fourier mode samples within the survey volume, and
\begin{eqnarray}
    P_\mathrm{shot}(\vec{k}) &=& \frac{e^{k_\parallel^2 \sigma_r^2}}{n_{\mathrm{gal}}}
\end{eqnarray}
is the shot noise contribution, which stems from the finite number density of galaxies. The exponential factor is a window function that accounts for damping of line-of-sight modes because of the finite redshift accuracy of the galaxy survey, $\sigma_r = c \sigma_z / H(z)$, where $\sigma_z$ is the galaxy redshift uncertainty, for example for a typical spectroscopic ($\sigma_z\approx0.001)$, grism ($\sigma_z\approx0.01$), or photometric survey ($\sigma_z\approx0.1$).
We note that individual narrow-band filters typically cover only $\Delta z\sim 0.1-0.2$, which is smaller than the assumed line-of-sight depth of our survey. Therefore, the predictions presented below for $\sigma_z=0.1$ correspond to the case in which either multiple narrow-band filters or measurements are grouped together, or a prism covers a redshift interval of at least $\Delta z=0.36$. 

\paragraph{The binned cross-power spectrum uncertainty $\sigma_\mathrm{21,gal}(\hat{k})$:}
Following Eq.~\ref{eq:sigma_cross}, we evaluate the uncertainty of the cross-power spectrum for each individual Fourier mode $\vec{k}$. Although the Universe is statistically isotropic on large scales, observational uncertainties (and redshift space distortions, which we omit in this analysis) affect line-of-sight and transverse modes differently. Therefore, we calculate both the 2D 21cm--galaxy cross-power spectrum uncertainties $P_\mathrm{21,gal}(k_\parallel, k_\perp)$, and the spherically averaged (one-dimensional) cross-power spectrum uncertainties, $P_\mathrm{21,gal}(k)$. To do this, we obtained the total uncertainty $\sigma_\mathrm{21,gal}(\hat{k})$ by inverse-variance summation over all contributing Fourier modes $\vec{k}$ within each $\hat{k}$ bin while accounting for the number of independent pointings;
\begin{equation}
    \sigma_\mathrm{21,gal}(\hat{k}) = \left[\sum_i^{N_m(\hat{k})} \frac{1}{\sigma^2_\mathrm{21,gal}(\vec{k}_i)} \right]^{-1/2} \times\ \left(\frac{\Omega'}{\Omega_\mathrm{survey}}\right)^{1/2} . 
\end{equation}
Here $\hat{k}$ denotes either the 2D pair $(k_\parallel, k_\perp)$ for the cylindrical two-dimensional case or the scalar $k = \sqrt{k_\parallel^2 + k_\perp^2}$ for the spherically averaged one-dimensional case. The number of modes, $N_m(\hat{k})$, in each $\hat{k}$ bin depends on the array configuration of the interferometer (i.e. number of independent baselines), the assumed integration time (i.e. number of independent observations), and the removal of modes contaminated by spectrally smooth foregrounds at low $k_\parallel$ that couple into higher $k_\parallel$ modes through the chromatic response of the instrument (known as the foreground wedge), and modes outside of the survey footprint. 
The angular footprint of the survey, defined by the solid angle of the FoV, $\Omega_\mathrm{survey}$, sets the minimum accessible transverse mode, $k_\perp^\mathrm{min}=\frac{2\pi}{X} \frac{1}{\sqrt{\Omega_\mathrm{survey}}}$, while the instrumental spatial resolution, given by the shortest baselines, defines the maximum, $k_\perp^\mathrm{max}$.
Analogously, along the line of sight, the total bandwidth $B$ sets the maximum extent $L_\parallel = YB$, which in turn determines the minimum accessible mode, $k_\parallel^\mathrm{min} = \frac{2\pi}{YB}$, while the spectral resolution of the instrument sets the maximum $k_\parallel^\mathrm{max} = n_\mathrm{channels} k_\parallel^\mathrm{min}$ with $n_\mathrm{channel}=164$ for SKA-Low1. Since the fundamental Fourier mode spacing is $2\pi/L_\parallel$, we adopt bin sizes of $\frac{2\pi}{YB}$ in both the $k_\parallel$ and $k_\perp$ directions.
The number of pointings is determined by $\Omega_\mathrm{survey}$ and the effective beam solid angle, $\Omega'$. Using $\Omega'$ instead of the beam solid angle $\Omega_p$ accounts for the overlap between individual pointings caused by the flux density gradually decaying towards the sidelobes of the beam. We follow \citet{Pober2014} and assume $\Omega'=2\Omega_p$. 

\begin{figure}
\centering
\includegraphics[width=\hsize]{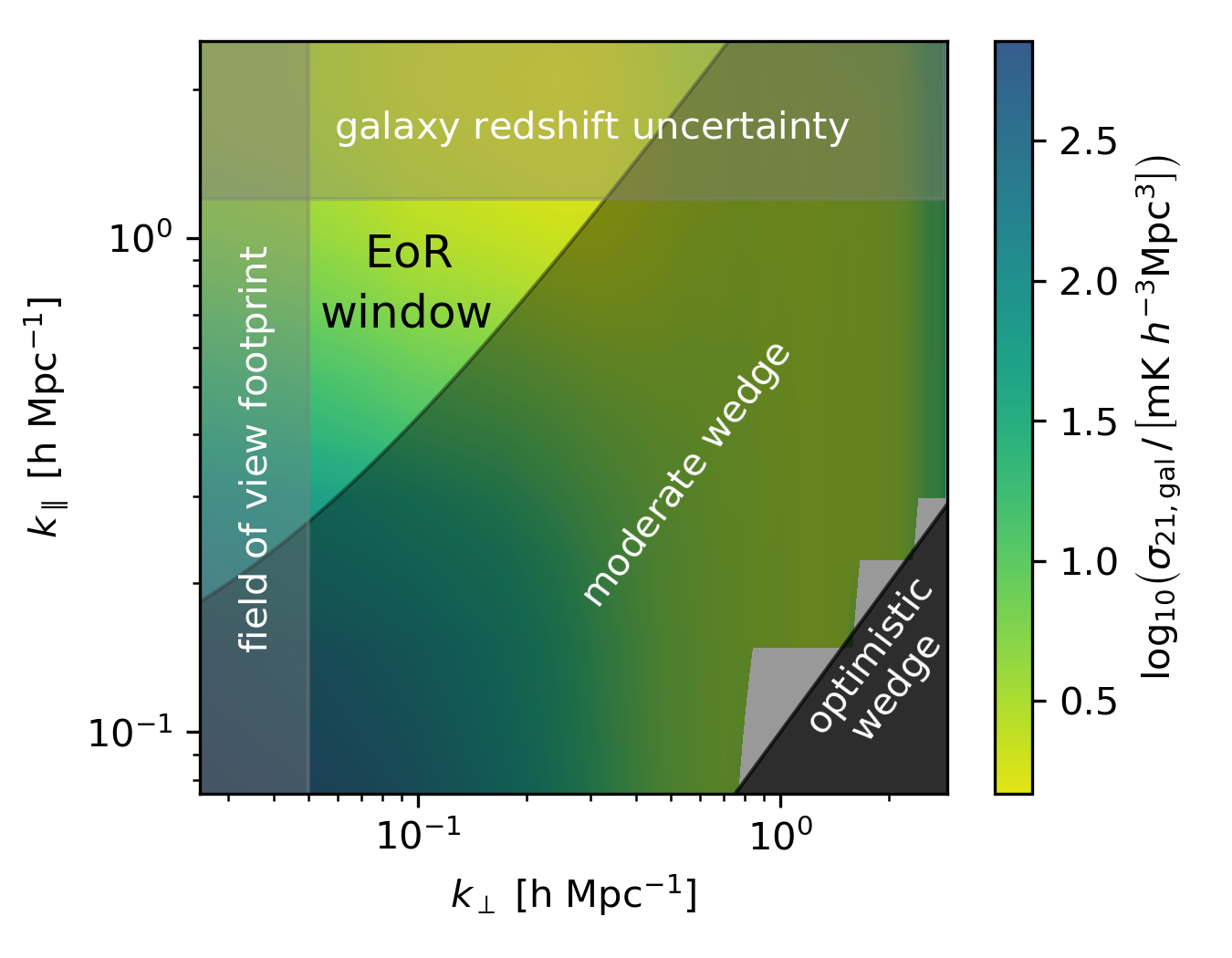}
  \caption{Two-dimensional 21cm--galaxy cross-power uncertainties, $\sigma_\mathrm{21,gal}(k_\parallel, k_\perp)$, (see Sec.~\ref{sec_error}). The accessible EoR window for cross-correlations depends on (1) 21cm foreground assumption and (2) galaxy survey redshift uncertainty.}
 \label{fig_SNR_k}
\end{figure}

In Fig.~\ref{fig_SNR_k}, we show the 2D 21cm--galaxy cross-power uncertainties, $\sigma_\mathrm{21,gal}(k_\parallel, k_\perp)$, for a galaxy survey detecting Ly$\alpha$ emitters down to $L_\alpha^\mathrm{min}=10^{42}$erg~s$^{-1}$ over a FoV of $100$~deg$^2$ with the SKA-Low1 AA4 array. We indicate three regions in the $k_\parallel$-$k_\perp$ space where modes are effectively inaccessible: (1) the foreground wedge, caused by the chromatic response of the instrument to spectrally smooth foregrounds and described by
\begin{eqnarray}
    k_{\parallel} &\leq& 
    \begin{cases}
        \left( \frac{X}{Y} \frac{1}{\nu} \right)\, k_{\perp} + 0.1 h\, {\rm Mpc}^{-1}, & {\rm moderate~ wedge} \\
        \left( \frac{X}{Y} \frac{1}{\nu} \right)\, k_{\perp} \sin\left( 0.61\, d \right), & {\rm optimistic~ wedge}
    \end{cases}
\end{eqnarray}
where $\nu = \frac{c}{\lambda_{21}(z)}$ is the observed frequency and $d$ the dish size in units of the redshifted 21cm line wavelength, $\lambda_{21}(z)$, (2) the redshift damping region, where fluctuations on scales smaller than the galaxy redshift uncertainty cannot be detected, and (3) the field-of-view footprint, where fluctuations larger than the survey FoV cannot be measured.
In practice, we assume modes in the marked foreground wedge regions (shown as shaded regions) to have infinite variance, meaning they are excluded when computing $\sigma_\mathrm{21,gal}(\hat{k})$. At high $k_\parallel$ values, $\sigma_\mathrm{21,gal}$ also diverges to infinity due to the finite precision of galaxy redshifts (shown as the upper shaded region). This leaves only intermediate $k_\parallel$ and small to intermediate $k_\perp$ modes, the so-called EoR window, as contributors to the measurable cross-power signal. 
As a note, for a galaxy survey covering a narrow redshift slice of $\Delta z = 0.1$, the corresponding bandwidth ($B = 2.2$~MHz) gives $k_\parallel^\mathrm{min} = \frac{2\pi}{YB} = 0.27\, h$~Mpc$^{-1}$, which exceeds the maximum $k_\parallel$ set by a photometric-like redshift uncertainty ($\sigma_z=0.1$, $k_\parallel^\mathrm{max} = \frac{2\pi}{\sigma_r}=0.18\,h$~Mpc$^{-1}$), leaving no EoR window.

The colour scale in Fig.~\ref{fig_SNR_k} illustrates how the cross-power uncertainty $\sigma_\mathrm{21,gal}$ scales within the EoR window in the optimistic foreground scenario (including the moderate foreground wedge). At the largest transverse scales (smallest $k_\perp$ values), the finite FoV limits the number of accessible modes -- for example, a FoV of $1$~deg$^2$, smaller as compared to the $100$~deg$^2$ assumed here, corresponds to $k_\perp^\mathrm{min}\sim0.06h$~Mpc$^{-1}$. At large scales both in parallel and transverse direction (small $k$ values), the uncertainty is dominated by cosmic variance, as only a few independent modes fit into the survey volume. As such, the uncertainty follows the 21cm--galaxy cross-power spectrum, decreasing in amplitude towards higher $k$ values. 
However, towards smaller scales (larger $k$ values), the uncertainty rises as the number of long baselines decreases needed for accessing larger $k_\perp$ values, increasing the thermal noise in the 21cm measurement, and as the shot noise from the finite galaxy number density begins to dominate at large $k_\parallel$ values. 
Moreover, we find an increase in $\sigma_\mathrm{21,gal}$ at larger $k_\perp$ values arises where the SKA-Low1 AA4 array has fewer baselines, which coincide with modes lying within the moderate foreground wedge. As a result, the optimal detection regime for the 21cm-galaxy cross-power spectrum lies typically at intermediate scales, $k\sim 0.1$--$0.5\,h~\mathrm{Mpc}^{-1}$, depending on the FoV, galaxy redshift precision and number density.

These trends already hint at that the detailed survey design and foreground treatment crucially shape $\sigma_\mathrm{21,gal}(k_\parallel, k_\perp)$, and thus the achievable S/N of the 21cm--galaxy cross-power spectrum. Therefore, we summarise below how survey parameters and the choice of the foreground wedge affect the accessible modes and resulting uncertainty. 
\begin{itemize}
    \item Field of view: A larger FoV increases the total number of accessible modes, reducing $\sigma_\mathrm{21,gal}(k_\parallel, k_\perp)$ due to lower $k_\perp$ modes being accessible.
    \item Galaxy redshift uncertainty, $\sigma_z$: Increasing $\sigma_z$ damps larger line-of-sight modes, lowering the maximum $k_\parallel$ value of the EoR window and raising $\sigma_\mathrm{21,gal}(k_\parallel, k_\perp)$ at high $k_\parallel$ values.
    \item Limiting Ly$\alpha$ luminosity, $L_\alpha^\mathrm{min}$: A higher $L_\alpha^\mathrm{min}$ decreases the galaxy number density, which can further reduce the maximum $k_\parallel$ value of the EoR window if it falls below the galaxy redshift uncertainty limit, effectively increasing the shot noise term.
    \item Foreground wedge assumption: A more optimistic foreground wedge assumption (recovering all modes larger than the beam) retains more modes with finite $\sigma_\mathrm{21,gal}(k_\parallel, k_\perp)$ values than a moderate foreground wedge (excluding modes below the horizon plus a $0.1\,h$~Mpc$^{-1}$ buffer).
\end{itemize}

\section{Signal detection and signal-to-noise analysis}
\label{sec_SNR}

We used the aforementioned {\sc astraeus} simulations to assess which galaxy survey designs, characterised by the FoV, redshift uncertainty and limiting Ly$\alpha$ luminosity, are best suited to detect the cross-power spectrum between the 21cm signal and LAEs during reionisation. For this purpose, we examine the scale-dependent (Section~\ref{subsec_survey_param}) as well as the total (Section~\ref{subsec_SNR_surveys}) S/N of the 21cm--LAE cross-power spectra. We also test whether neglecting the auto-power spectra in the cross-power uncertainty calculations -- allowing the uncertainties to be derived from analytical fitting functions of the 21cm--LAE cross-correlation function \citep{Hutter2023b} -- yields results comparable to those obtained when including the auto-power spectra (see Appendix~\ref{sec_cosmic_variance}).

For each set of survey parameters, we derive the scale-dependent S/N from the spherically averaged cross-power spectra and uncertainties as
\begin{eqnarray}
    (\mathrm{S/N})_k& = & \frac{P_\mathrm{21,gal}(k)}{\sigma_\mathrm{21,gal}(k)},
    \label{eq:SNR_k}
\end{eqnarray} 
and the total S/N from the square root of the sum of the quadratures of the S/N in each $(k_\parallel, k_\perp)$ bin,
\begin{eqnarray}
    \mathrm{S/N} &=& \sqrt{\sum_i^{N_\mathrm{bins}} \frac{P_\mathrm{21,gal}^2(k_\parallel, k_\perp)}{\sigma_\mathrm{21,gal}^2(k_\parallel, k_\perp)}}, 
    \label{eq:SNR}
\end{eqnarray}
with $N_\mathrm{bins}$ being the number of contributing $(k_\parallel, k_\perp)$ bins. As noted in the previous section, both, $\sigma_\mathrm{21,gal}(k)$ and $\sigma_\mathrm{21,gal}(k_\parallel, k_\perp)$ already account for the number of modes $N_m(k)$ and $N_m(k_\parallel, k_\perp)$ measured by SKA-Low1. 
We further note that we have derived the following results for both reionisation scenarios, {\sc mhdec} and {\sc mhinc}, simulated with {\sc astraeus}. We adopt the {\sc mhdec} simulation (faint galaxies driving reionisation) as our default to examine $(\mathrm{S/N})_k$ and S/N dependence on survey parameters, noting differences in the {\sc mhinc} results where relevant.

\citet{Fronenberg2024} noted that when the cross-power spectrum passes through zero, the S/N defined as $P_{\rm 21,gal}/\sigma_{\rm 21,gal}$ becomes unreliable for signal detection. In our analysis, however, the zero-crossing of the cross-power spectrum occurs at $k$ values dominated by galaxy shot noise (see Fig. \ref{fig_crossps_Lmin}, \ref{fig_SNR_k_mod}, and \ref{fig_SNR_k_opt}). Therefore, we do not expect this effect to significantly affect the interpretation of our reported S/N values.

\subsection{$k$-dependent signal-to-noise and survey parameter dependence}
\label{subsec_survey_param}

\begin{figure}
\centering
\includegraphics[width=\hsize]{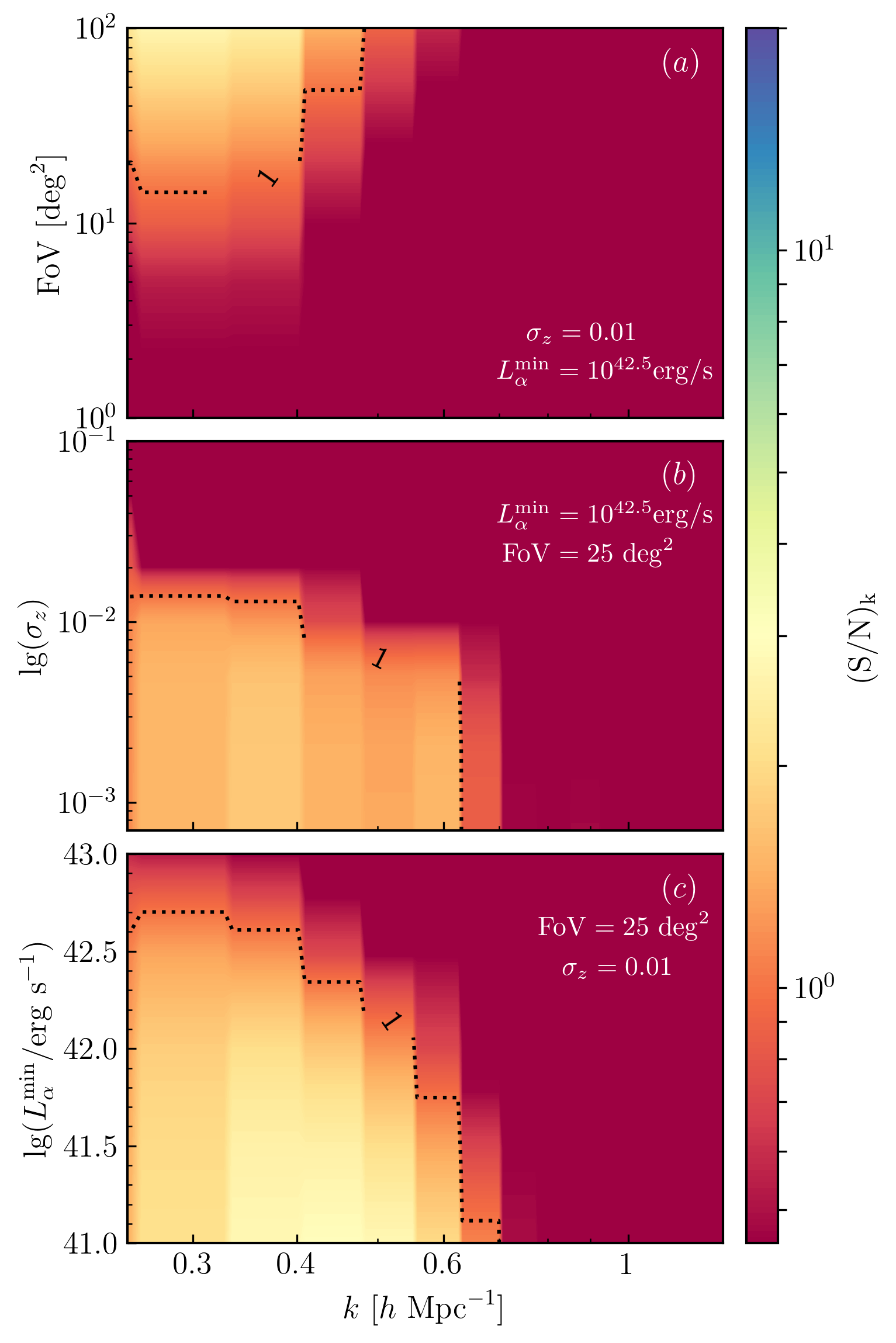}
  \caption{Signal-to-noise ratio of the 21cm--LAE cross-power spectrum $P_\mathrm{21,LAE}(k)$ as a function of wavenumber $k$ for different observational configurations. The top, middle, and bottom panels correspond to varying FoV, galaxy redshift uncertainty, $\sigma_z$, and minimum Lyman-$\alpha$, luminosity $L_\alpha^\mathrm{min}$, respectively. The 21cm signal noise is computed assuming the SKA-Low1 AA4 antenna layout and a moderate foreground model.}
     \label{fig_SNR_k_mod}
\end{figure}

\begin{figure}
\centering
\includegraphics[width=\hsize]{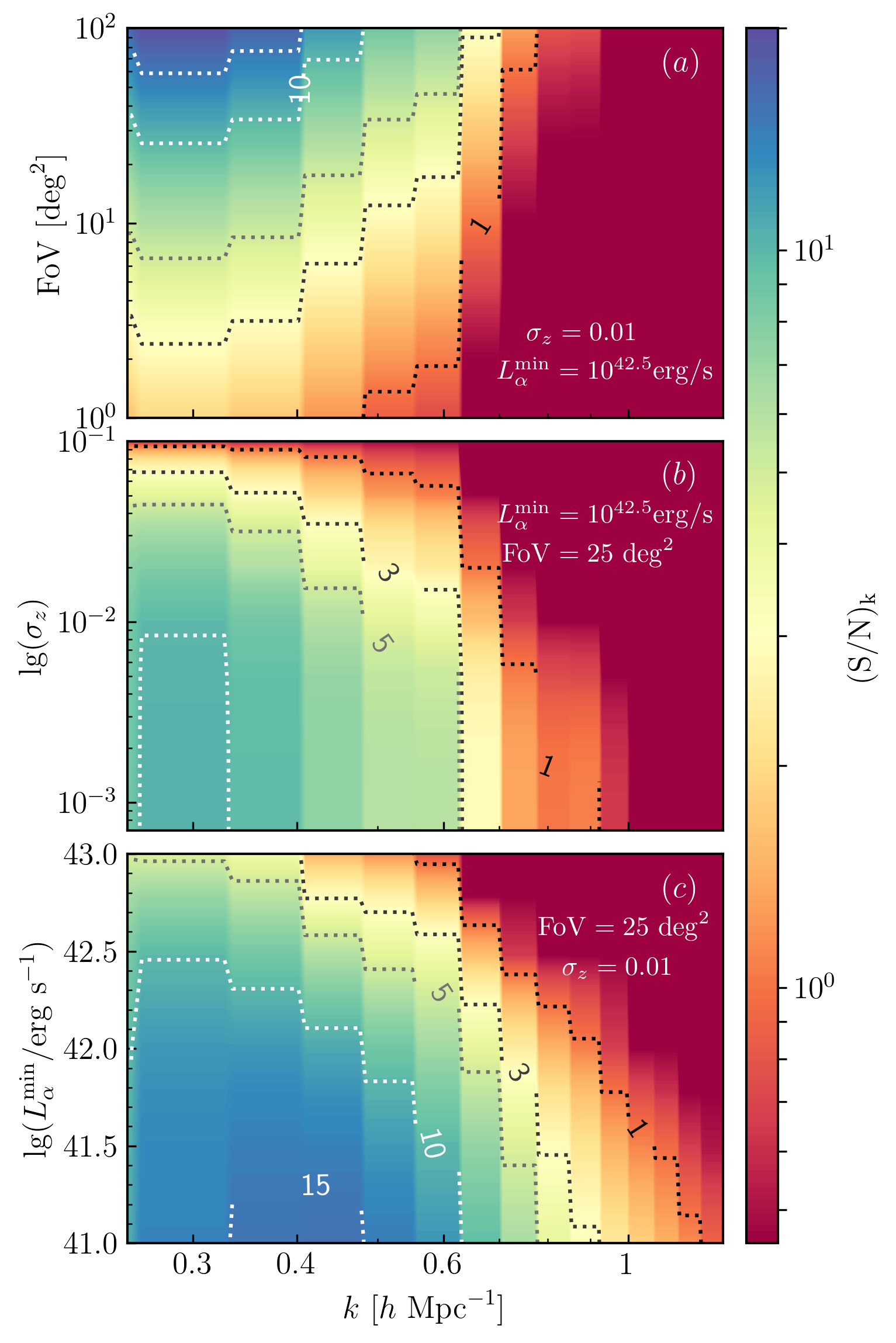}
  \caption{Signal-to-noise ratio of the 21cm-LAE cross-power spectrum $P_\mathrm{21,LAE}(k)$ as a function of wavenumber $k$ for different observational configurations. The top, middle, and bottom panels correspond to varying FoV, galaxy redshift uncertainty, $\sigma_z$, and minimum Lyman-$\alpha$ luminosity, $L_\alpha^\mathrm{min}$, respectively. The 21cm signal noise is computed assuming the SKA-Low1 AA4 antenna layout and an optimistic foreground model.}
     \label{fig_SNR_k_opt}
\end{figure}

As shown in Section~\ref{sec_error}, the cross-power spectrum uncertainty has a strong $k$-dependence. This $k$-dependence is critical for assessing how sensitive the 21cm--LAE cross-power spectrum is to variations in the ionisation morphology and how well it can distinguish between different reionisation scenarios.
To identify which $k$ modes provide the highest S/N, we show in Fig.~\ref{fig_SNR_k_mod} and \ref{fig_SNR_k_opt} how $(\mathrm{S/N})_k$ varies with the FoV (top panel), the galaxy survey redshift uncertainty $\sigma_z$ (middle panel) and minimum detectable Ly$\alpha$ luminosity $L_\alpha^\mathrm{min}$ (bottom panel) for moderate and optimistic foreground wedge assumptions at $z=7$, respectively. We choose to display this redshift here, as we already have a large number of, and expect more, LAE detection as well large FoVs covered at $z\sim7$. 
In general, from large to small scales (small to large $k$ values), $(\mathrm{S/N})_k$ rises first and peaks around scales of $k\sim0.3-0.5\, h$~Mpc$^{-1}$ before it drops sharply. 

On large scales, $k\lesssim0.4\,h$~Mpc$^{-1}$, the 21cm-LAE cross-power spectrum uncertainties are dominated by cosmic variance (see Fig.~\ref{fig_SNR_cosmicVariance_k}), with the relative impact set by the FoV and the low $k_\parallel$ modes lost to foregrounds.
Therefore, as shown in the top panels of Fig.~\ref{fig_SNR_k_mod} and \ref{fig_SNR_k_opt}, increasing the FoV, which allows sampling of more independent large-scale modes, decreases the uncertainties and increases the S/N. Since the thermal noise is negligible compared to the intrinsic 21cm signal variance on these scales ($P_\mathrm{noise} \ll P_\mathrm{21}$), the dominant terms in Eq.~\ref{eq:sigma_cross} are $P_{21} P_\mathrm{gal}$ and $P_{21} P_\mathrm{shot}$. Among these, $P_{21} P_\mathrm{shot}$ drives the increase in $(\mathrm{S/N})_k$ when more galaxies are detected (reducing $P_\mathrm{shot}$ towards lower $L_\alpha^\mathrm{min}$ values in bottom panels) and their positions are better resolved (reducing $P_\mathrm{shot}$ towards lower $\sigma_z$ values in middle panels).

On intermediate scales, $k\simeq 0.4 - 0.7\,h$~Mpc$^{-1}$, the interferometer sensitivity declines as the number of long baselines decreases. Consequently, the thermal noise becomes significant, causing all terms in Eq.~\ref{eq:sigma_cross} to contribute. As a result, the cross-power spectrum uncertainties, and thus $(\mathrm{S/N})_k$, become strongly dependent on all survey parameters: A larger FoV, sampling more modes, decreases the thermal noise contribution, while the shot noise contribution is mitigated by increasing the galaxy number density through lowering $L_\alpha^\mathrm{min}$, or reducing the redshift uncertainty of the galaxies $\sigma_z$.

\begin{figure*}
\resizebox{\hsize}{!}
        {\includegraphics{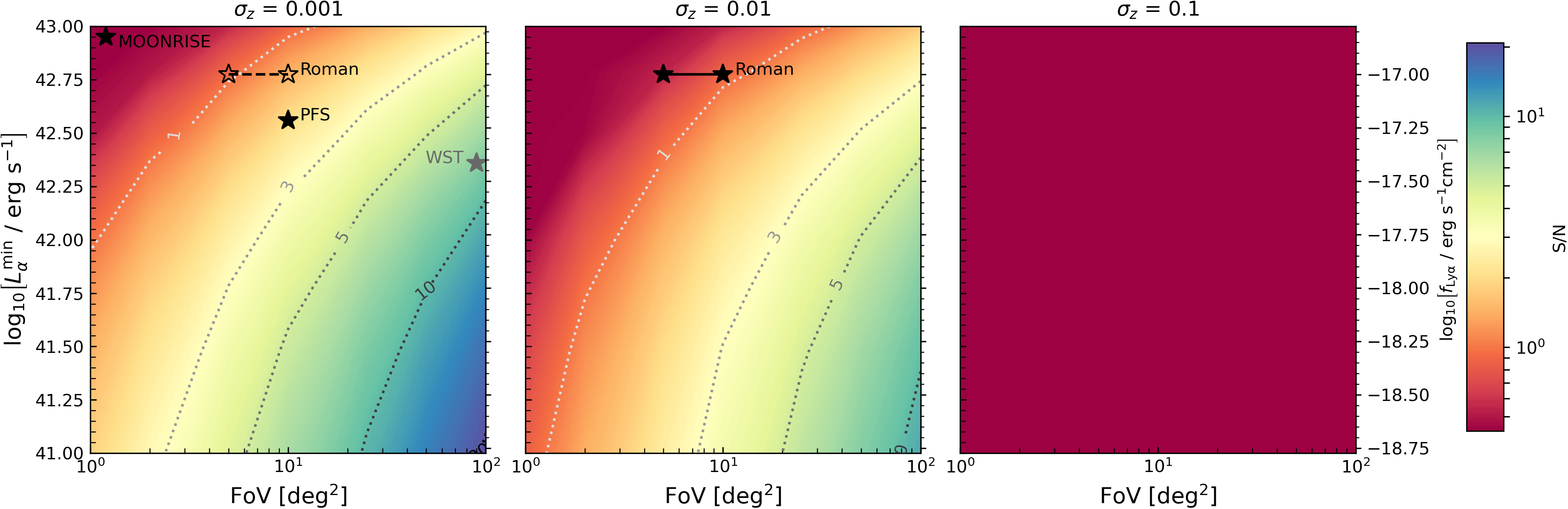}}
  \caption{Total S/N of the 21cm--LAE cross-power spectrum in the {\sc mhdec} reionisation scenario as a function of survey FoV and the minimum Lyman-$\alpha$ luminosity, $L_\alpha^\mathrm{min}$, shown for spectroscopic (left), grism (centre) and photometric-like (right) surveys at $z=7$. The 21cm signal noise was computed assuming the SKA-Low1 AA4 antenna layout and a moderate foreground model. Black stars mark potential LAE surveys at $z\simeq7$ that could be cross-correlated with SKA 21cm data (see also Tab.~\ref{tab_surveys}).}
     \label{fig_totalSNR_paramspace_MHDEC_mod}
\end{figure*}

\begin{figure*}
\resizebox{\hsize}{!}
        {\includegraphics{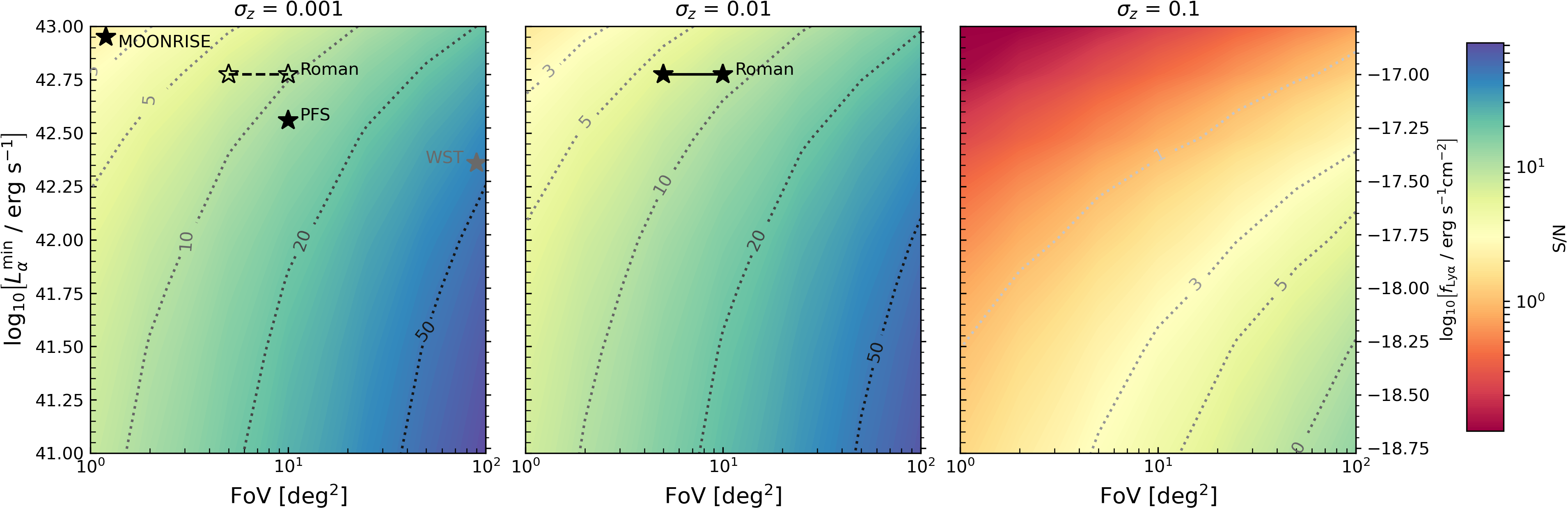}}
  \caption{Total S/N of the 21cm--LAE cross-power spectrum in the {\sc mhdec} reionisation scenario as a function of survey FoV and the minimum Lyman-$\alpha$ luminosity, $L_\alpha^\mathrm{min}$, shown for spectroscopic (left), grism (centre) and photometric-like (right) surveys at $z=7$. The 21cm signal noise is computed assuming the SKA-Low1 AA4 antenna layout and an optimistic foreground model. Black and grey stars mark potential LAE surveys at $z\simeq7$ that could be cross-correlated with SKA 21cm data (see also Tab.~\ref{tab_surveys}).}
     \label{fig_totalSNR_paramspace_MHDEC_opt}
\end{figure*}

The exact scale below which the $(\mathrm{S/N})_k$ becomes shot-noise-dominated and drops abruptly, $k_\mathrm{shot\, noise}$, depends on all galaxy survey parameters. For a fixed shot noise contribution (i.e. constant $L_\alpha^\mathrm{min}$ and $\sigma_z$), reducing the relative thermal noise contribution ($P_\mathrm{noise}$) by enlarging the galaxy survey FoV and thus the number of detectable modes shifts $k_\mathrm{shot\, noise}$ to larger $k$ values (smaller physical scales). Conversely, for a fixed galaxy survey FoV and thus thermal noise contribution, raising the shot noise contribution pushes $k_\mathrm{shot\, noise}$ to smaller $k$ values. Both the minimum detectable Ly$\alpha$ luminosity and the galaxy redshift uncertainty affect the shot noise. Lowering $L_\alpha^\mathrm{min}$ increases the galaxy number density, thereby reducing the overall shot noise amplitude, while reducing $\sigma_z$ allows access to smaller line-of-sight scales, pushing the shot-noise-dominated regime to smaller physical scales. 
As a consequence, once the average distance between the detected galaxies, $\sim 1/n_\mathrm{gal}^{1/3}$, becomes larger than the line-of-sight scale corresponding to $\sigma_z$, a further reduction in $\sigma_z$ no longer improves the measurement. In this regime, the shot-noise is instead determined by the galaxy number density, such that $k_\mathrm{shot\, noise}(L_\alpha^\mathrm{min}) > k_\mathrm{shot\, noise}(\sigma_z)$. For instance, from the middle panels in Fig.~\ref{fig_SNR_k_mod} and \ref{fig_SNR_k_opt}, we see that both in a moderate and optimistic foreground scenario for $L_\alpha^\mathrm{min} = 10^{42.5}$erg~s$^{-1}$, which corresponds to a number density of $n_\mathrm{gal}\simeq10^{-3}$Mpc$^{-3}$ at $z=7$ in our simulations, using spectroscopic surveys ($\sigma_z\simeq0.001$) instead of grism surveys ($\sigma_z\simeq0.01$) does not significantly improve $(\mathrm{S/N})_k$.

Beyond these galaxy survey parameters, the assumed 21cm foreground wedge, i.e. which modes are recoverable, also strongly impacts the S/N levels. This dependence is evident when comparing the moderate wedge (Fig.~\ref{fig_SNR_k_mod}) to optimistic wedge model (Fig.~\ref{fig_SNR_k_opt}). Extending the EoR window to larger $k_\perp$ values allows more modes $N_\mathrm{bins}=\sum N_m(k_\parallel, k_\perp)$ to be sampled, which reduces the overall uncertainty and raises the overall S/N level. As a result, $k_\mathrm{shot\, noise}$ shifts to smaller physical scales.

Finally, it is evident from Fig.~\ref{fig_SNR_k_mod} and \ref{fig_SNR_k_opt} that the S/N at small physical scales, where the sign change in the 21cm--LAE cross-power traces the typical size of the ionised regions around the selected galaxy population (see Section~\ref{subsec_eor_galaxy}), is too low to distinguish between reionisation scenarios. However, the high S/N at large to intermediate scales offers a promising avenue, as these modes are sensitive to the mean 21cm differential brightness temperature, $\langle \delta T_b \rangle$, and thus trace the mean neutral hydrogen gas density, $\langle \delta_\mathrm{HI} \rangle$. The evolution of $\langle \delta_\mathrm{HI} \rangle$ with the mean ionisation fraction, $\langle\chi_\mathrm{HII} \rangle$, or redshift, depends on how the ionisation fronts propagate through the cosmic web: the more strongly the ionising emissivity correlates with the underlying gas density -- such as in scenarios where fainter, less massive galaxies dominate the ionising photon budget -- the lower $\langle \delta_\mathrm{HI} \rangle$ is at a fixed $\langle \chi_\mathrm{HII} \rangle$.
Measuring the 21cm--LAE cross-power at large scales with forthcoming SKA observations and galaxy surveys (e.g. Roman, PFS on Subaru, or proposed missions such as REX) will constrain the average neutral gas density and hence the ionisation history and morphology. However, breaking the degeneracy between ionisation history and morphology will require measurements across multiple scales.

\subsection{Total signal-to-noise and survey parameter dependence}
\label{subsec_SNR_surveys}

Next, to assess which survey parameter combinations are best for obtaining a significant 21cm--LAE cross-power signal, we show in Fig.~\ref{fig_totalSNR_paramspace_MHDEC_mod} and \ref{fig_totalSNR_paramspace_MHDEC_opt} the total S/N as defined in Eq.~\ref{eq:SNR} as a function of $L_\alpha^\mathrm{min}$ and FoV for spectroscopic ($\sigma_z=0.001$), grism ($\sigma_z=0.01$) and photometric-like ($\sigma_z=0.1$) surveys, assuming a moderate and optimistic foreground wedge, respectively.

Firstly, and as already discussed in Section~\ref{subsec_survey_param}, we see that the S/N increases with survey volume (larger FoV) and as more galaxies are probed (lower $L_\alpha^\mathrm{min}$). This results from either an overall increase in the number of modes or an broadening of the EoR window to smaller physical scales along the line-of-sight. The constant S/N contours in the figures further indicate that lowering $L_\alpha^\mathrm{min}$ is less effective at increasing the S/N than expanding the FoV, particularly as galaxy positions become better resolved (lower $\sigma_z$). This trend arises because enlarging the FoV increases the total number of modes, providing more large-scale high S/N measurements, whereas decreasing the shot noise adds comparably fewer high $k_\parallel$ modes. 

Secondly, it is evident that improving the positional accuracy of galaxies (lower $\sigma_z$) increases the S/N. This occurs because a lower $\sigma_z$  broadens the EoR window to include larger $k_\parallel$ modes. However, we note that the increase in the S/N slows down and, for high S/N values, saturates as $\sigma_z$ decreases further (c.f. contours for $\sigma_z=0.01$ and $0.001$ and optimistic foreground wedge in Fig.~\ref{fig_totalSNR_paramspace_MHDEC_opt}). This behaviour arises because, for a fixed $L_\alpha^\mathrm{min}$, the number of additional high $k_\parallel$ modes gained by decreasing $\sigma_z$ drops once the line-of-sight scale corresponding to $\sigma_z$ becomes smaller than the average distance between the selected galaxies. 

Finally, we note that achieving the same S/N for detecting the 21cm--LAE cross-power spectrum in a bright-galaxy-driven reionisation scenario ({\sc mhinc}) as in a faint-galaxy-driven scenario ({\sc mhdec}, shown in the figures) requires either a FoV roughly twice as large or the detection of Ly$\alpha$ emission from galaxies about three times fainter (see Appendix \ref{sec_snr_mhinc}).

\begin{table*}
    \centering
    \caption{Characteristics of the forthcoming Ly$\alpha$-emitting galaxy surveys.}
    \label{tab_surveys}
    \begin{tabular*}{\textwidth}{ccccc}
        \hline
        Instrument & Type & FoV  & Ly$\alpha$ flux limit & Comment \\
        \hline
        \hline
        PFS & spectroscopic & 10~deg$^2$ & $10^{-17.2}$ erg s$^{-1}$ cm$^{-2}$ & potential survey \\
        &&&& \citet{Greene2022} for flux limit\\
        MOONRISE & spectroscopic & 1~deg$^2$ & $10^{-16.8}$ erg s$^{-1}$ cm$^{-2}$ & \citet{Maiolino2020} \\
        Widefield Spectroscopic Telescope & spectroscopic &  100~deg$^2$ & $10^{-17.4}$  erg s$^{-1}$ cm$^{-2}$ & potential survey with future \\ &&&& NIR extension to $z\sim 7$ \\
        Roman & grism ($\sigma_z\sim0.003$) & 5-10~deg$^2$ & $10^{-17}$ erg s$^{-1}$ cm$^{-2}$ & potential survey \\
        \hline
    \end{tabular*}
\end{table*}

\paragraph{Galaxy surveys suitable for detecting the 21cm--LAE cross-power at $z\sim7$:} 
From Fig.~\ref{fig_totalSNR_paramspace_MHDEC_mod}, we find that a statistically significant detection (S/N~$\gtrsim3$) with photometric-like galaxy surveys is not feasible if the 21cm signal cannot be recovered beyond the horizon (moderate wedge scenario). 
Thus, in the case of a restricted EoR window, only large-area (FoV$\gtrsim10-20$~deg$^2$), medium-deep ($L_\alpha^\mathrm{min}\sim10^{42.5}$erg~s$^{-1}$) spectroscopic or grism surveys can achieve S/N~$\simeq3$. 
Such surveys are within reach of next-generation facilities such as the PFS on Subaru, which can follow up Roman prism observations, or the proposed Widefield Spectroscopic Telescope (WST). 
In a more optimistic scenario, where the 21cm signal can be recovered within the horizon and beyond the instrumental beam, a detection (S/N~$\gtrsim3$) becomes feasible also with wide ($\sim100\,$deg$^2$), medium-deep ($L_\alpha^\mathrm{min}\sim10^{42.3}$erg~s$^{-1}$) photometric-like surveys. For grism and spectroscopic galaxy surveys, the requirements become substantially less demanding: smaller areas ($\sim2\,$deg$^2$) and shallower depths ($L_\alpha^\mathrm{min}\sim10^{42.8}$erg~s$^{-1}$) could still yield a detection with S/N~$\sim3$. Such surveys could be conducted with PFS or Roman, whereas the smaller-area MOONRISE survey -- the main MOONS Extragalactic GTO program \citep{Maiolino2020} -- does not cover sufficient area to obtain a robust signal.

In summary, detecting most effectively the 21cm--LAE cross-power spectrum at $z=7$ requires large-area, medium-deep to deep grism or spectroscopic galaxy surveys. However, if the 21cm signal can be recovered beyond the horizon, the required galaxy surveys can instead be smaller and shallower.

\section{Survey strategies for distinguishing reionisation scenarios}
\label{sec_reionisation_scenarios}

\begin{figure*}
\resizebox{\hsize}{!}
        {\includegraphics{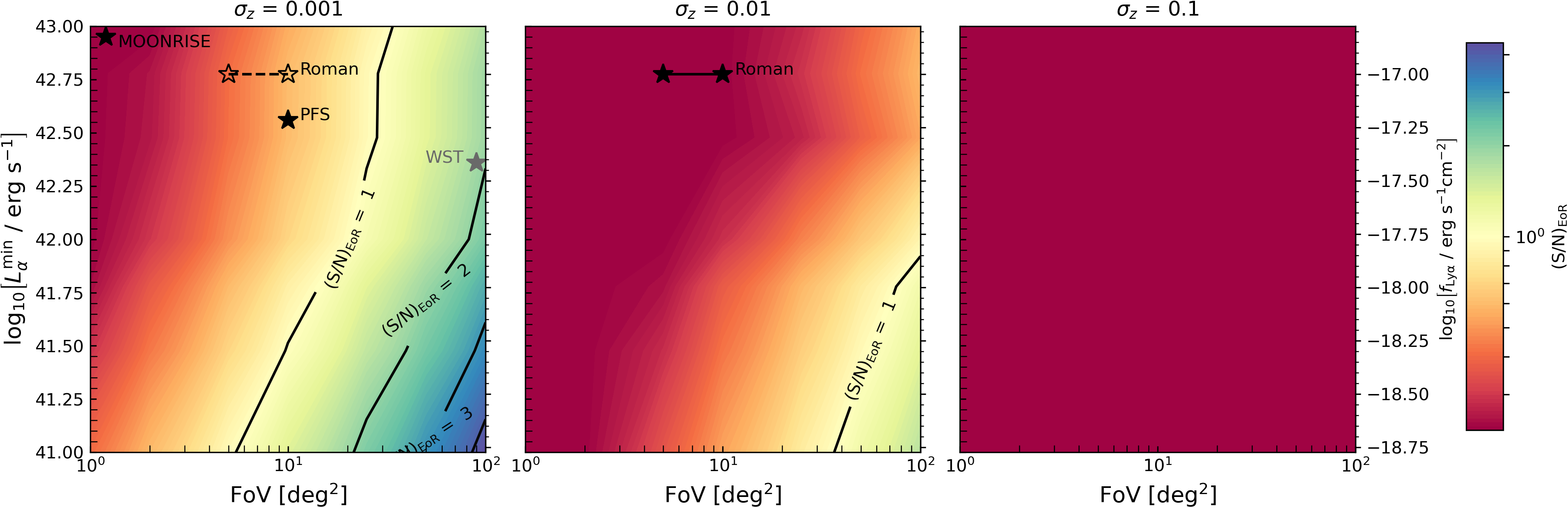}}
  \caption{Total S/N (black lines and coloured contours) for distinguishing the {\sc mhdec} and {\sc mhinc} reionisation scenarios using the $z=7$ 21cm--LAE cross-power spectra as a function of the survey FoV and the minimum Lyman-$\alpha$ luminosity, $L_\alpha^\mathrm{min}$, shown for spectroscopic (left), grism (centre) and photometric-like (right) surveys. The 21cm signal noise is computed assuming the SKA-Low1 AA4 antenna layout and a moderate foreground model. Black and grey stars mark potential LAE surveys at $z\simeq7$ that could be cross-correlated with SKA 21cm data (see also Tab.~\ref{tab_surveys}).}
     \label{fig_totalSNR_paramspace_reionisation_scenarios_mod}
\end{figure*}

\begin{figure*}
\resizebox{\hsize}{!}
        {\includegraphics{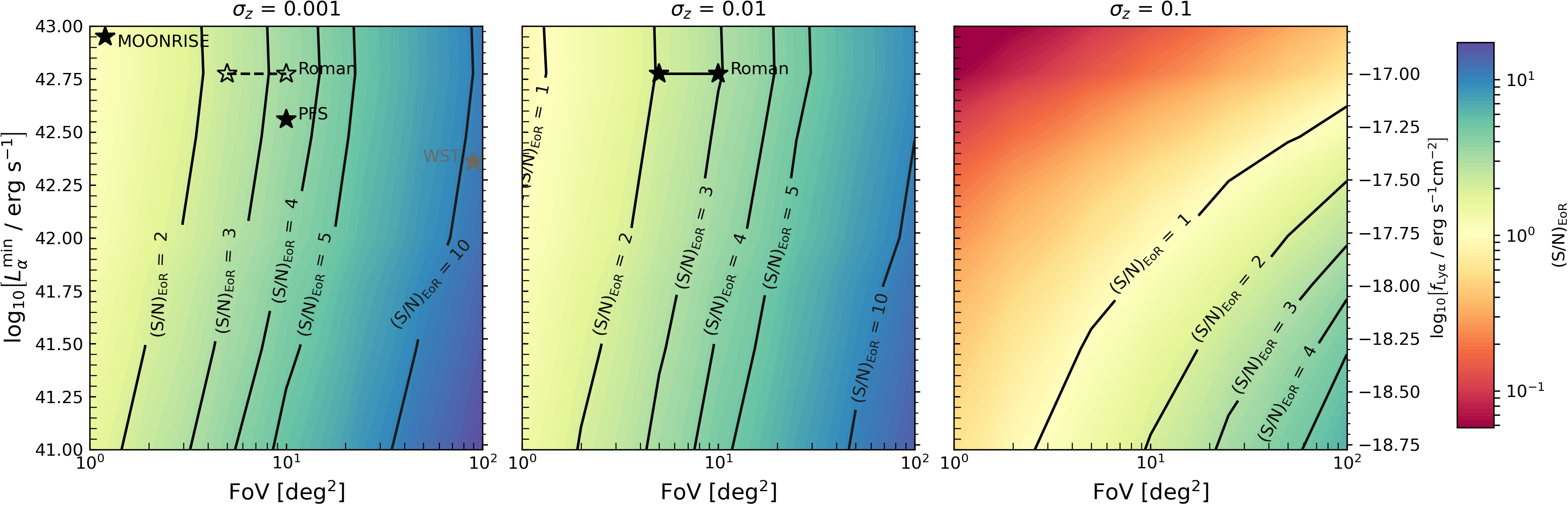}}
  \caption{Total S/N (black lines and coloured contours) for distinguishing the {\sc mhdec} and {\sc mhinc} reionisation scenarios using the $z=7$ 21cm--LAE cross-power spectra as a function of the survey FoV and the minimum Lyman-$\alpha$ luminosity, $L_\alpha^\mathrm{min}$, shown for spectroscopic (left), grism (centre) and photometric-like (right) surveys. The 21cm signal noise is computed assuming the SKA-Low1 AA4 antenna layout and an optimistic foreground model. Black and grey stars mark potential LAE surveys at $z\simeq7$ that could be cross-correlated with SKA 21cm data (see also Tab.~\ref{tab_surveys}).}
     \label{fig_totalSNR_paramspace_reionisation_scenarios_opt}
\end{figure*}

Having defined the sets of survey parameters that enable a statistical detection of the 21cm--LAE cross-power spectrum at $z=7$, we now turn to an equally interesting question: For which survey configurations would a cross-power spectrum measurement be able to distinguish between two contrasting reionisation scenarios -- one in which faint galaxies drive reionisation ({\sc mhdec}) or one in which bright galaxies ({\sc mhinc}) are the primary drivers?
To answer this question, we compare the cross-power spectra predicted by the two scenarios in each ($k_\parallel, k_\perp$) bin, perform a $\chi^2$ analysis, and define the corresponding S/N as 
\begin{equation}
    \mathrm{(S/N)_{EoR}} = \sqrt{ \sum_{(k_\parallel, k_\perp)}^{N_\mathrm{bins}} \frac{\left[ P_\mathrm{21,gal, [1]}(k_\parallel, k_\perp) - P_\mathrm{21,gal, [2]}(k_\parallel, k_\perp) \right]^2}{\frac{1}{2} \left[ \sigma_\mathrm{21,gal, [1]}^2(k_\parallel, k_\perp) + \sigma_\mathrm{21,gal, [2]}^2(k_\parallel, k_\perp) \right]} }, 
\end{equation}
where the subscripts $[1]$ and $[2]$ represent the different scenarios. The resulting $\mathrm{(S/N)_{EoR}}$ at $z=7$ are shown for the parameter space of FoV, $L_\alpha^\mathrm{min}$, and $\sigma_z$ and the moderate and optimistic wedge models in Fig.~\ref{fig_totalSNR_paramspace_reionisation_scenarios_mod} and \ref{fig_totalSNR_paramspace_reionisation_scenarios_opt}.

Comparing these figures, it is evident -- similarly as before -- that we obtain far lower $\mathrm{(S/N)_{EoR}}$ values for the moderate than the optimistic wedge model. Again, for the moderate wedge model (Fig.~\ref{fig_totalSNR_paramspace_reionisation_scenarios_mod}), because photometric-like ($\sigma_z=0.1$) surveys cannot obtain a statistically significant detection of the cross-power spectrum, they likewise cannot separate the two contrasting reionisation scenarios. Even with grism and spectroscopic surveys, a substantial push towards large ($\mathrm{FoV}\gtrsim50$~deg$^2$) and deep ($L_\alpha^\mathrm{min}\lesssim10^{42.0}$erg~s$^{-1}$) surveys would be required to differentiate between scenarios ($\mathrm{(S/N)_{EoR}}\gtrsim1-3$). However, even such grism surveys could only marginally distinguish between them ($\mathrm{(S/N)_{EoR}}\simeq1$). The reason for this low distinguishing power is the smaller EoR window that hardly includes large-scale modes ($k\lesssim0.3\, h$~Mpc$^{-1}$) that trace the negative peak of the cross-power spectrum and are the least affected by galaxy shot noise. As we have outlined in Section~\ref{sec_signal}, the amplitude of this peak traces the average neutral hydrogen gas density, which, even at fixed global (volume-averaged) ionisation fractions, depends on the ionisation morphology. In short, the cross-power spectrum characteristic with the most constraining power is hardly sampled.
For this reason, extending the EoR window to smaller $k_\parallel$ and $k_\perp$ values, as it is the case in the optimistic wedge model (Fig.~\ref{fig_totalSNR_paramspace_reionisation_scenarios_opt}), causes a significant jump in the $\mathrm{(S/N)_{EoR}}$ values. Now medium-large (FoV$\simeq5-10$~deg$^2$), shallow ($L_\alpha^\mathrm{min}\simeq10^{42.8}$erg~s$^{-1}$) grism and spectroscopic surveys could distinguish between contrasting reionisation scenarios ($\mathrm{(S/N)_{EoR}}>3-5$), and even large (FoV$\sim50$~deg$^2$), deep ($L_\alpha^\mathrm{min}\simeq10^{42}$erg~s$^{-1}$) photometric-like surveys would reach significant distinguishing power ($\mathrm{(S/N)_{EoR}}\gtrsim3$).

For the spectroscopic and grism surveys, we see that the S/N to differentiate between contrasting reionisation scenarios depends more on the FoV than the galaxy number density. The reason for this behaviour arises from the fact that the large-scale modes, containing most of the constraining power, are hardly affected by the shot noise that decreases with rising galaxy number density. We see this imbalance for the optimistic wedge model more than for the moderate one, because the larger EoR window allows sampling more large-scale modes.

\begin{figure*}
\resizebox{\hsize}{!}
        {\includegraphics{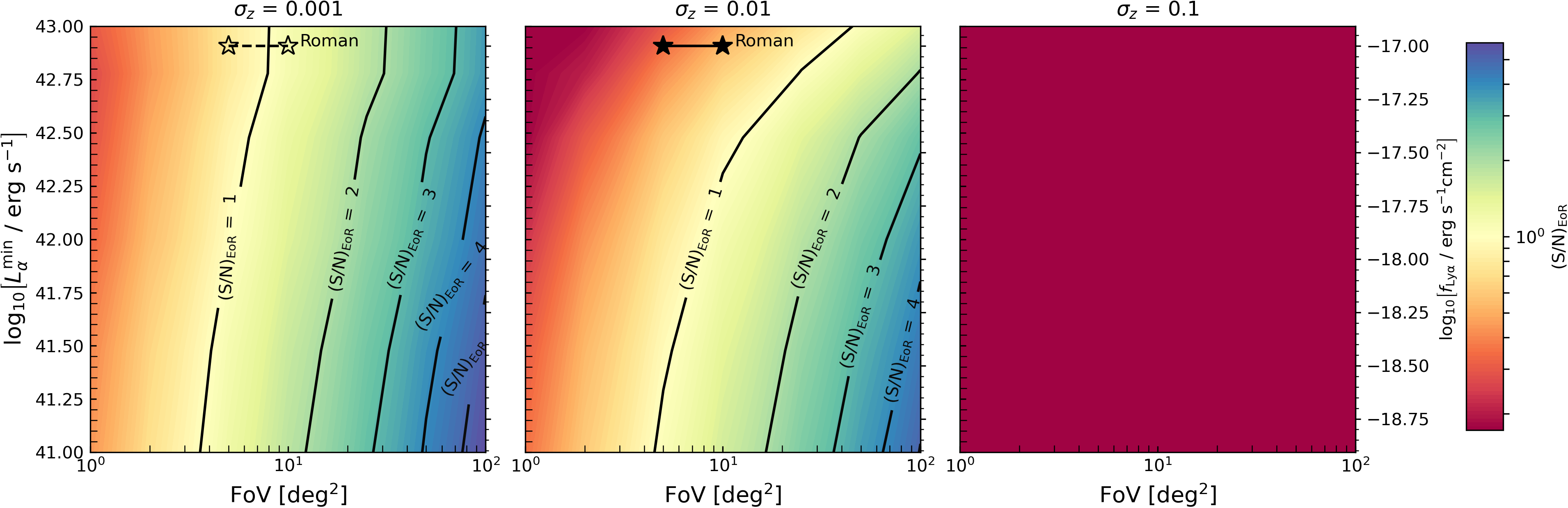}}
  \caption{Total S/N (black lines and coloured contours) for distinguishing the {\sc mhdec} and {\sc mhinc} reionisation scenarios using the $z=8$ 21cm--LAE cross-power spectra as a function of the survey FoV and the minimum Lyman-$\alpha$ luminosity, $L_\alpha^\mathrm{min}$, shown for spectroscopic (left), grism (centre) and photometric-like (right) surveys. The 21cm signal noise is computed assuming the SKA-Low1 AA4 antenna layout and a moderate foreground model.}    \label{fig_totalSNR_paramspace_reionisation_scenarios_mod_z8}
\end{figure*}

Finally, it is important to note that our two contrasting reionisation scenarios exhibit only minor differences in the global ionisation fraction, $\langle \chi_\mathrm{HII} \rangle$. Thus, the differences in the cross-power spectra arise mainly from the scenario's different ionisation morphologies. At higher (or lower) redshifts, however, the scenario's global ionisation fractions diverge more significantly, leading to different amounts of the maximum negative amplitude of their cross-power spectra being increased (decreased) and their peaks being shifted to smaller (larger) physical scales (see Fig.~\ref{fig_crossps_Lmin}). Given the scales sampled within the EoR window, this change in relevant $k$ scales implies that different reionisation scenarios can be distinguished with even smaller, shallower surveys. This is illustrated in Fig.~\ref{fig_totalSNR_paramspace_reionisation_scenarios_mod_z8} which shows the results for the moderate wedge model at $z=8$. However, at higher redshifts, the increase in $\mathrm{(S/N)_{EoR}}$ for a fixed set of survey parameters arises not only from the difference in $\langle\chi_\mathrm{HII}\rangle$ but also from the shift of the negative cross-power spectrum peak to larger $k$ values, where more modes are available and thus better sampled.

This implies two main conclusions: Firstly, as already found in previous works \citep[e.g.][]{Hutter2017, Heneka2017, Heneka2021, Weinberger2020, Hutter2023b}, the 21cm--LAE cross-power spectrum can be used to constrain the global ionisation fraction, $\langle\chi_\mathrm{HII}\rangle$. Secondly, moving to higher redshifts and lower $\langle \chi_\mathrm{HII} \rangle$ values facilitates both the detection of the cross-power spectrum\footnote{We have derived the total S/N at $z=8$ for the {\sc mhdec} reionisation scenario (analogous to Fig.~\ref{fig_totalSNR_paramspace_MHDEC_mod}), yielding slightly higher values across all combinations of survey parameters.} and the distinction between contrasting reionisation scenarios, although for fixed luminosity thresholds the galaxy number density decreases. At $z>8$, however, spin temperature fluctuations alter the trends of the cross-power spectrum, and detectable galaxies become even sparser. 
For observations around $z\sim7$, our results suggest that recovering large-scale modes beyond the horizon will be crucial for increasing the sensitivity needed to constrain reionisation.
Importantly, even with a tenfold reduction in SKA-Low1 integration time, the S/N and $\mathrm{(S/N)_{EoR}}$ decrease by only $\sim30$\% (see Appendix~\ref{sec_snr_100h}), allowing 21cm observations to cover a substantially larger FoV.

An 8~MHz bandwidth corresponds to $\Delta z \simeq 0.36$, $0.46$, and $0.56$ at $z = 7$, $8$, and $9$, respectively. Between $z = 7$ and $8$, combining two independent redshift slices would only yield a modest increase in the total S/N.

\section{Conclusions}
\label{sec_conclusions}

We have investigated which 21cm signal and galaxy survey characteristics are required (1) to detect the 21cm--LAE cross-power spectrum at $z=7$ and $8$, and (2) to distinguish between different reionisation scenarios and constrain the ionisation morphology. Our analysis focuses on two contrasting scenarios explored with the {\sc astraeus} framework in \citet{Hutter2023a}: one in which faint galaxies drive reionisation, and another where bright galaxies dominate the ionising photon budget. We consider three key survey parameters: the common FoV of both surveys, the minimum observed Ly$\alpha$ luminosity ($L_\alpha^\mathrm{min}$), and the redshift uncertainty of the galaxy survey ($\sigma_z$). In addition, we determine the impact of the 21cm foreground scenario. By exploring a wide range of survey configurations -- $\mathrm{FoV}=1-100$~deg$^2$, $L_\alpha^\mathrm{min}=10^{41}-10^{43}$erg~s$^{-1}$, and $\sigma_z=0.1$, $0.01$, $0.001$ (corresponding to photometric-like, grism and spectroscopic surveys, respectively) -- we reach the following conclusions:

\begin{itemize}
    \item The negative peak amplitude of the 21cm--galaxy cross-power spectrum traces the average neutral hydrogen gas density, while its scale reflects the sizes of ionised regions around the probed galaxy population. Both features are sensitive to the IGM's ionisation history and morphology.
    \item The S/N for detecting the 21cm--galaxy cross-power spectrum increases with larger survey FoV, higher galaxy number density (corresponding to a lower limiting Ly$\alpha$ luminosity, $L_\alpha^\mathrm{min}$), and improved redshift accuracy of the galaxy survey, $\sigma_z$. Expanding the FoV is generally more effective -- especially at lower $\sigma_z$ -- as it adds more large-scale modes, whereas reducing $L_\alpha^\mathrm{min}$ mainly lowers shot noise and contributes fewer additional high-$k_\parallel$ modes. Additionally, for fixed $L_\alpha^\mathrm{min}$, the S/N gain from improving $\sigma_z$ saturates once the redshift resolution surpasses the typical galaxy separation, because hardly more high-$k_\parallel$ modes are added.
    \item The foreground wedge model affects the S/N for detecting the 21cm-LAE cross-power spectrum: 
    \begin{itemize}
        \item Moderate foreground wedge (modes below the horizon and a $0.1, h$ Mpc buffer excluded): detection is not possible with photometric-like surveys. A detection ($\mathrm{S/N} \gtrsim 3$) becomes possible with medium-deep ($L_\alpha^\mathrm{min} \simeq 10^{42.5}$ erg s$^{-1}$), wide-area ($\mathrm{FoV} \gtrsim 50$ deg$^2$) grism or medium-area ($\mathrm{FoV} \gtrsim 20$ deg$^2$) spectroscopic surveys.
        \item Optimistic foreground wedge (modes recoverable down to the beam limit): a robust detection ($\mathrm{S/N} \geq 3$) is feasible with deep ($L_\alpha^\mathrm{min} \simeq 10^{42.3}$ erg s$^{-1}$), wide-area ($\mathrm{FoV} \gtrsim 80$ deg$^2$) photometric-like surveys. Grism and spectroscopic surveys are less sensitive to $L_\alpha^\mathrm{min}$, enabling detections even with shallow, small-area surveys ($\mathrm{FoV} \simeq 2-3$ deg$^2$).
    \end{itemize}
    \item Distinguishing contrasting reionisation scenarios via 21cm-LAE cross-power spectra is harder when only the ionisation morphology differs (as at $z=7$) than when both the global ionisation fraction and morphology differ (as at $z=8$). Consequently, the S/N for model discrimination ($\mathrm{(S/N)_{EoR}}$) is lower at $z=7$. Distinguishing our two reionisation scenarios at $z=7$ is possible for the following survey characteristics:
    \begin{itemize}
        \item Moderate foreground wedge: a marginal detection ($\mathrm{S/N} \gtrsim 1$) may be achieved with deep ($L_\alpha^\mathrm{min} \simeq 10^{42}$ erg s$^{-1}$), wide-area ($\mathrm{FoV} \gtrsim 100$ deg$^2$) grism surveys. A more robust detection ($\mathrm{(S/N)_{EoR}} \gtrsim 3$) requires deeper and equally wide spectroscopic surveys.
        \item Optimistic foreground wedge: a robust detection ($\mathrm{(S/N)_{EoR}}\geq3$) is possible with deep ($L_\alpha^\mathrm{min}\simeq10^{42}$erg~s$^{-1}$), wide-area ($\mathrm{FoV}\gtrsim100$~deg$^2$) photometric-like surveys. Grism and spectroscopic surveys can still reach significant $\mathrm{(S/N)_{EoR}}$ values with shallower and more compact configurations ($\mathrm{FoV} \gtrsim 10$ deg$^2$).
    \end{itemize}
    \item S/N for both detection and model discrimination is maximised when the EoR window covers the large-scale modes of the negative cross-power spectrum peak. As reionisation progresses, this peak shifts to larger physical scales, which in our reionisation scenarios are no longer fully sampled by the EoR window at $z\sim7$ under a moderate foreground wedge.
\end{itemize}

We briefly list the main caveats. 
Firstly, our 21cm--LAE cross-power spectrum predictions are limited to redshifts where the IGM is fully heated ($z\lesssim8$, \citealt{HERA2023}), as the simulations employed here do not yet model spin temperature fluctuations -- a feature we plan to incorporate in future work. At higher redshifts, and depending on the reionisation scenario, the large-scale 21cm--galaxy cross-power spectrum is expected to exhibit a sign-flip, indicating that the IGM is still undergoing heating \citep{Heneka2020,Moriwaki2024}. However, detectability at higher redshifts may be challenging due to lower galaxy number densities, despite the stronger 21cm signal and recent JWST observations suggesting higher-than-expected galaxy abundances.

Secondly, our analysis does not account for the light cone effect; that is, we assume the average ionisation fraction, $\langle \chi_\mathrm{HII} \rangle$, remains constant across the redshift interval $\Delta z = 0.36$. In reality, particularly around $z=7$, $\langle\chi_{\rm HII}\rangle$ can decrease by $0.1-0.15$ over this interval. Accounting for this effect would likely enhance the S/N, as differences in both the ionisation morphology and the ionisation history would contribute.

Thirdly, our analysis neglects redshift space distortions. These distortions enhance the large-scale cross-power spectrum and shift its peak to larger scales, as overdense regions are compressed and underdense regions stretched along the line of sight. While this boost could aid detection, the peak's shift to lower $k_\parallel$ may move it into the foreground wedge. The net impact on S/N is uncertain, but redshift space distortions may help distinguish between reionisation scenarios, as they are expected to more strongly affect cases where the ionisation field traces the density field, amplifying differences in the cross-power spectrum. 

Fourthly, although the 21cm and LAE fields are intrinsically non-Gaussian, we estimate cross-power spectrum uncertainties assuming Gaussian statistics. Empirical tests (App.~\ref{sec_covariances}) show that mode-mode correlations from thermal noise, redshift uncertainties, and survey geometry are negligible. If many independent modes contributing to a $k$ bin, \citet{Fronenberg2024} demonstrated that uncertainties are well approximated by a Gaussian distribution over several standard deviations. Non-Gaussian contributions that do not generate significant off-diagonal covariance (e.g. trispectrum or intra-bin mode coupling) are neglected and may modestly increase the true error bars.

In summary, detecting the 21cm--LAE cross-correlation signal depends critically on recovering 21cm modes near and below the horizon limit -- that is, within the foreground wedge. Under conservative assumptions (moderate foreground wedge), a detection will require large grism or spectroscopic galaxy surveys. However, advances in 21cm foreground cleaning techniques~\citep[e.g.][]{acharya2024, Bianco2021, Bianco2024, Bianco2025, GagnonHartman2021, Kennedy2024, Sabti2025} could enable foreground removal (optimistic foreground wedge), potentially allowing detections even with large photometric-like surveys.
Upcoming telescopes such as the PFS on Subaru, Roman or the proposed WST offer strong prospects for constraining the average IGM ionisation fraction, though probing the ionisation morphology would still require recovering 21cm modes hidden by foregrounds (i.e. assuming an optimistic foreground wedge). 
Achieving such constraints may require statistics sensitive to anisotropies, such as 21cm--galaxy wavelet cross-correlations or cross-bispectra, whose potential and observability we plan to explore in future works. In addition, non-Gaussian, machine-learning-based statistics~\citep{zhao2021, neutsch2022, prelogovic2023, saxena2023, schosser2024, ore2024}, explored for the 21cm signal, provide complementary avenues for improving cross-correlation measurements.

\section*{Data availability}

The code used to compute the 21cm-galaxy cross-power spectrum uncertainties is publicly available at \url{https://github.com/annehutter/21cm_gal_uncertainties}.

\begin{acknowledgements}
The authors thank the anonymous reviewer for their constructive and detailed comments. The authors also thank K. Kakiichi, S. Malhotra and J. Rhoads for discussions on future potential LAE surveys.
AH acknowledges support by the VILLUM FONDEN under grant 37459. The Cosmic Dawn Center (DAWN) is funded by the Danish National Research Foundation under grant DNRF140. 
CH’s work is funded by the Volkswagen Foundation. This work was supported by the DFG under Germany’s Excellence Strategy EXC 2181/1 - 390900948 The Heidelberg STRUCTURES Excellence Cluster.
\end{acknowledgements}

  \bibliographystyle{aa} 
  \bibliography{sources} 

\appendix 

\section{Cosmic variance contribution}
\label{sec_cosmic_variance}

\begin{figure}
\centering
\includegraphics[width=\hsize]{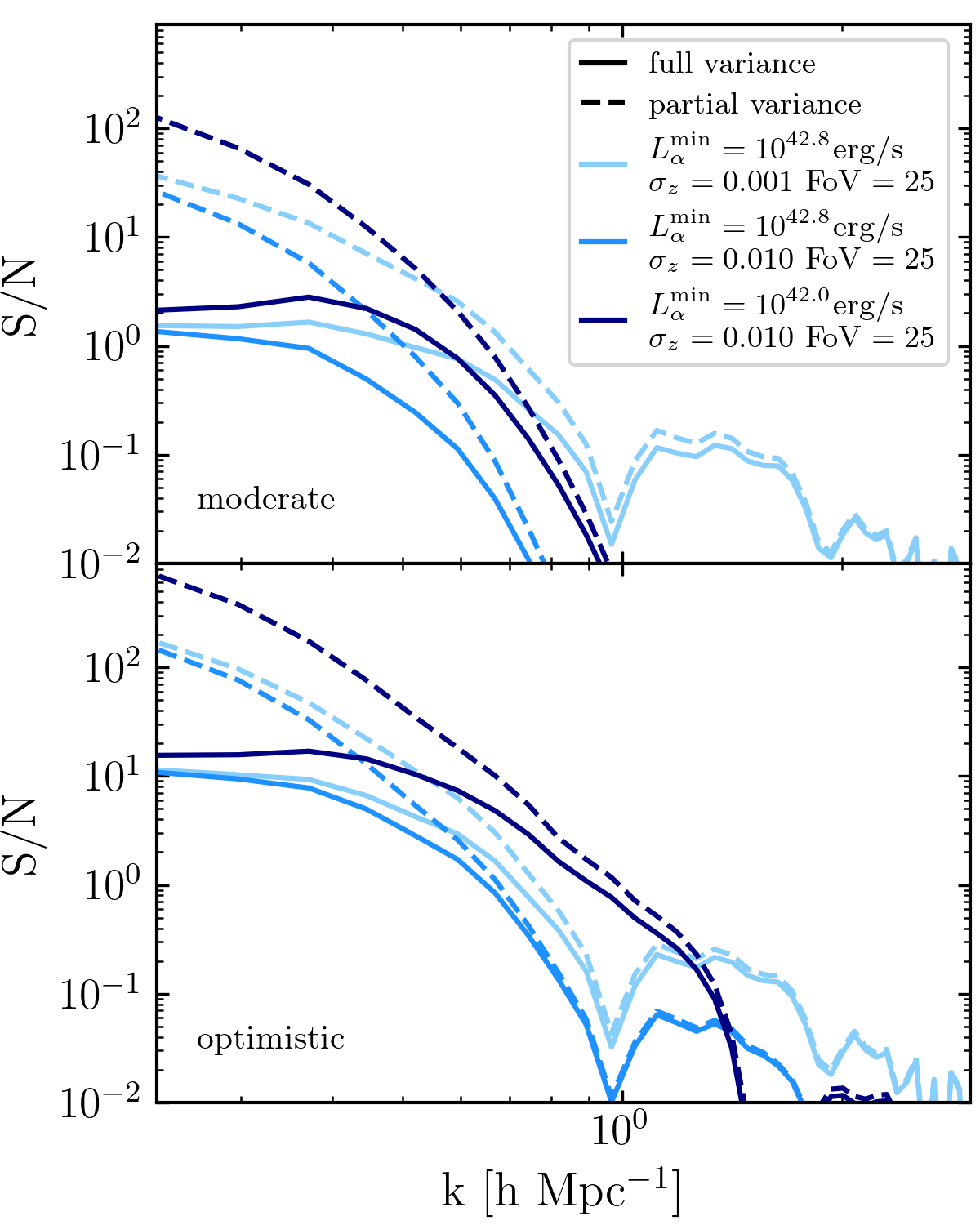}
  \caption{Signal-to-noise ratio of the $z=7$ {\sc mhdec} 21cm--LAE cross-power spectrum $P_\mathrm{21,LAE}(k)$ as a function of wavenumber $k$ assuming the moderate (top) and optimistic (bottom) foreground model for considering full uncertainties (solid lines) and uncertainties omitting the auto power spectra. Coloured lines show results for varying survey configurations.}
     \label{fig_SNR_cosmicVariance_k}
\end{figure}

Given that we developed an analytic fitting for function for the 21cm--LAE cross-correlation function in \citet{Hutter2023b}, in this section, we now assess whether omitting the auto-power spectrum contributions in the cross-power uncertainty calculation still provides a reasonable approximation at intermediate and small $k$ modes. 
For this purpose, Fig.~\ref{fig_SNR_cosmicVariance_k} shows the $k$-dependent $(\mathrm{S/N})_k$ computed with the full variance (solid lines) and with the partial variance (dashed lines), where $P_{21}=P_\mathrm{gal}=0$. We consider three different sets of survey parameters (coloured lines) and the two foreground models (panels). These parameter sets illustrate how $(\mathrm{S/N})_k$ changes when fainter galaxies are included (different $L_\alpha^\mathrm{min}$ values) and when galaxy positions are better resolved (different $\sigma_z$ values). 

We find that at large physical scales (small $k$), the uncertainties are dominated by the variances of the 21cm signal and galaxy number density fluctuations. On these scales, the partial-variance S/N is therefore higher than the full variance S/N. As the thermal noise becomes more important towards intermediate $k$ values, the difference between the two S/N estimates decreases, and at small physical scales (large $k$), the uncertainties are entirely dominated by galaxy shot noise.
Thus, for scales smaller than $r=2\pi/k$ with $k\sim0.7\,h^{-1}$~Mpc, the auto-power spectrum contribution to the uncertainty can be safely neglected. However, on larger scales -- which are most relevant for constraining reionisation -- the 21cm signal auto-power spectrum contribution remains significant.

\section{Correlation matrix}
\label{sec_covariances}

\begin{figure}
\centering
\includegraphics[width=\hsize]{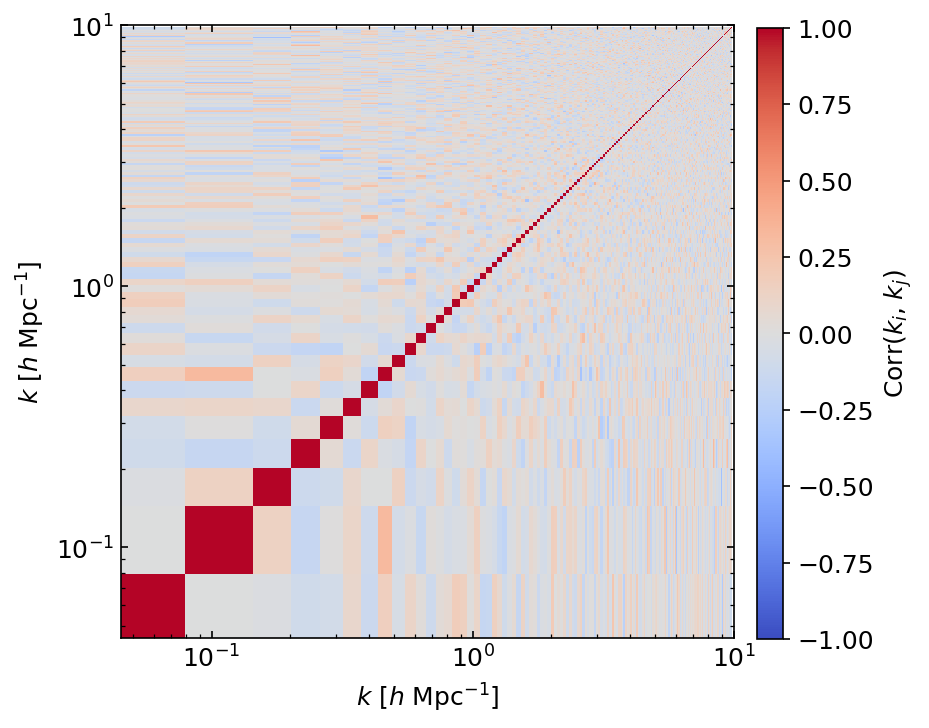}
  \caption{Correlation matrix for $100$ realisations of the $z=7$ {\sc mhdec} 21cm--LAE cross-power spectrum $P_\mathrm{21,LAE}(k)$. Assumed survey characteristics are $\sigma_z=0.01$, $L_\alpha^\mathrm{min}=10^{42.5}$erg~s$^{-1}$, $\mathrm{FoV}=25$~deg$^2$, the SKA-Low1 AA$\star$ array layout, and a moderate foreground wedge.}
     \label{fig_correlationMatrix}
\end{figure}

To assess whether different noise contributions are significantly correlated, we generate multiple realisations for the 21cm--LAE cross-power spectrum. For each realisation, galaxies in the survey are shifted within the assumed redshift uncertainty in real space and thermal noise, $P_\mathrm{noise}$, is added to the 21cm signal field in k-space. From the Fourier transformed fields, we compute the 21cm--galaxy cross-power spectrum accounting for the foreground wedge and survey dimensions. From $N=100$ realisations, the covariance matrix is derived as
\begin{eqnarray}
    \mathrm{Cov}(\vec{k}_i,\vec{k}_j) &=& \frac{\sum_{l=1}^{N} \left[ P_{l}(\vec{k}_i) -  \overline{P}(\vec{k}_i) \right] \left[ P_{l}(\vec{k}_j) - \overline{P}(\vec{k}_j) \right]}{N-1},
\end{eqnarray}
where $P(k)$ is the cross power spectrum. 
To assess how correlated different k-modes are, we calculate the correlation matrix as
\begin{eqnarray}
    \mathrm{Corr}(\vec{k}_i,\vec{k}_j) &=& \frac{\mathrm{Cov}(\vec{k}_i, \vec{k}_j)}{\sqrt{\mathrm{Cov}(\vec{k}_i, \vec{k}_i)~\mathrm{Cov}(\vec{k}_j, \vec{k}_j)}}.
\end{eqnarray}
Fig.~\ref{fig_correlationMatrix} shows $\mathrm{Corr}(\vec{k}_i,\vec{k}_j)$ for a wide-area medium-deep survey assuming a moderate 21cm foreground wedge. The off-diagonal elements are small compared to the diagonal terms and decrease with increasing number of realisations, becoming consistent with zero within the estimator noise. This indicates that, for the noise sources included here, mode–mode correlations in the cross-power spectrum covariance are negligible. It also suggest that non-Gaussian mode coupling induced by galaxy redhsifts uncertainties, thermal noise, survey geometry and wedge masking is negligible. Since we do not explicitly model the foregrounds, this analysis is not sensitive to any correlations they might induce.

\section{Signal-to-noise dependence in bright-galaxy-driven reionisation}
\label{sec_snr_mhinc}

\begin{figure*}
\centering
\includegraphics[width=\hsize]{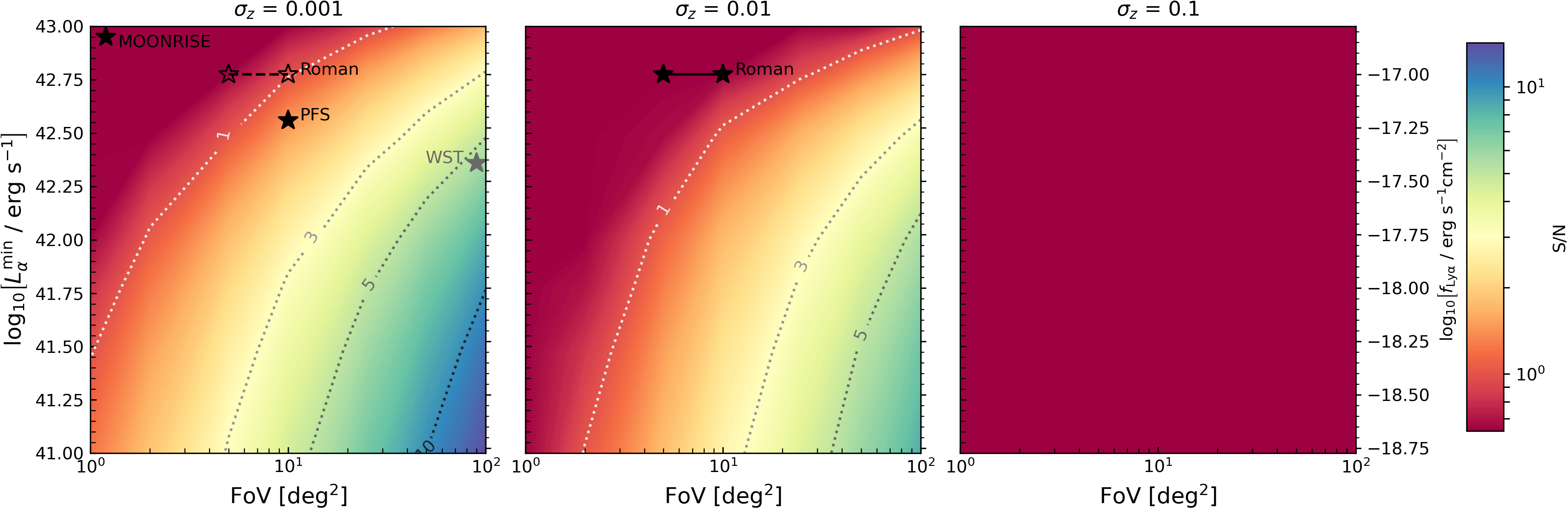}
    \caption{Total S/N of the 21cm--LAE cross-power spectrum in the {\sc mhinc} reionisation scenario as a function of survey FoV and the minimum Lyman-$\alpha$ luminosity, $L_\alpha^\mathrm{min}$, shown for spectroscopic (left), grism (centre) and photometric-like (right) surveys at $z=7$. The 21cm signal noise is computed assuming the SKA-Low1 AA4 antenna layout and a moderate foreground model. Black and grey stars mark potential LAE surveys at $z\simeq7$ that could be cross-correlated with SKA 21cm data (see also Tab.~\ref{tab_surveys}).}
    \label{fig_totalSNR_paramspace_MHINC_mod}
\end{figure*}

To assess how the S/N depends on the survey FoV, minimum Ly$\alpha$ luminosity, $L_\alpha^\mathrm{min}$, and redshift uncertainty, $\sigma_z$, under different reionisation scenarios, we show in Fig.~\ref{fig_totalSNR_paramspace_MHINC_mod} the S/N for detecting the 21cm-LAE cross-power spectrum for the {\sc mhinc} scenario, in which reionisation is driven by bright galaxies. Compared to the {\sc mhdec} scenario, achieving a similar S/N would require either a FoV roughly twice as large or the detection of Ly$\alpha$ emission from galaxies about three times fainter. 
The lower S/N for the same survey parameters arises because, in the {\sc mhinc} scenario, the ionisation field is more closely linked to massive, highly biased galaxies \citep[see][]{Hutter2023b}, which enhances cosmic variance effects relative to {\sc mhdec}.

\section{Signal-to-noise analysis for 100~h SKA-Low1 observations}
\label{sec_snr_100h}

\begin{figure*}
\resizebox{\hsize}{!}
        {\includegraphics{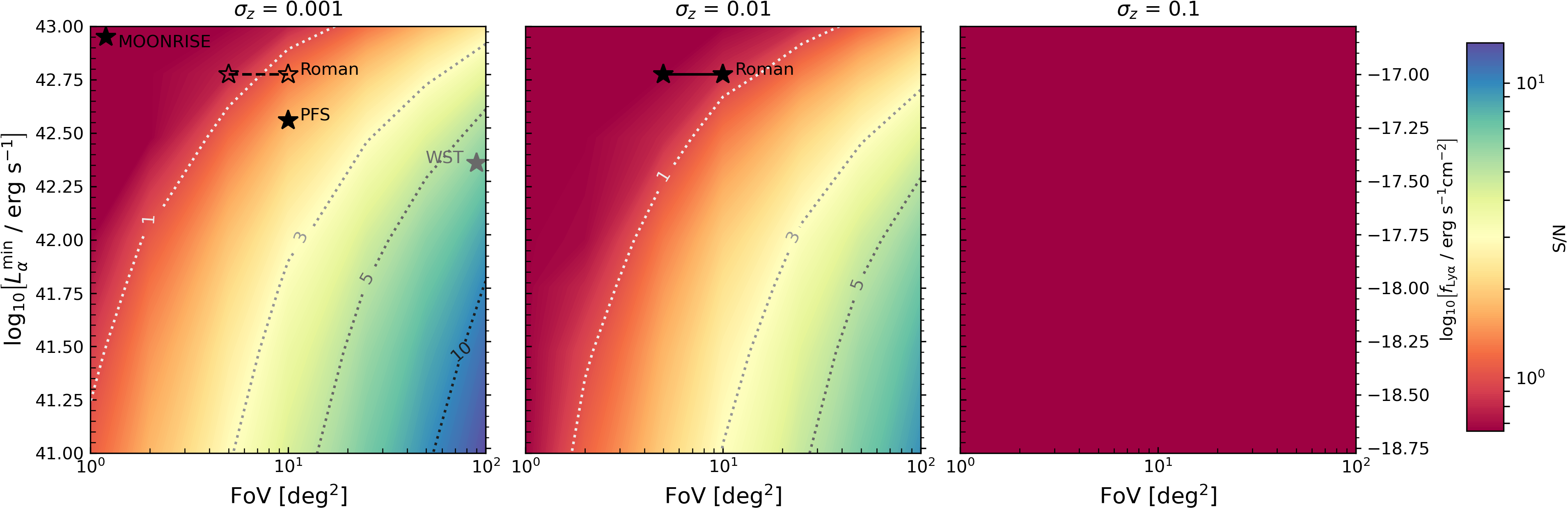}}
  \caption{Total S/N of the 21cm--LAE cross-power spectrum in the {\sc mhdec} reionisation scenario as a function of survey FoV and the minimum Lyman-$\alpha$ luminosity, $L_\alpha^\mathrm{min}$, shown for spectroscopic (left), grism (centre) and photometric-like (right) surveys at $z=7$. The 21cm signal noise is computed assuming the SKA-Low1 AA4 antenna layout, $\sim100~h$ integration time, and a moderate foreground model. Black and grey stars mark potential LAE surveys at $z\simeq7$ that could be cross-correlated with SKA 21cm data (see also Tab.~\ref{tab_surveys}).}
     \label{fig_totalSNR_paramspace_MHDEC_mod_100h}
\end{figure*}

\begin{figure*}
\resizebox{\hsize}{!}
        {\includegraphics{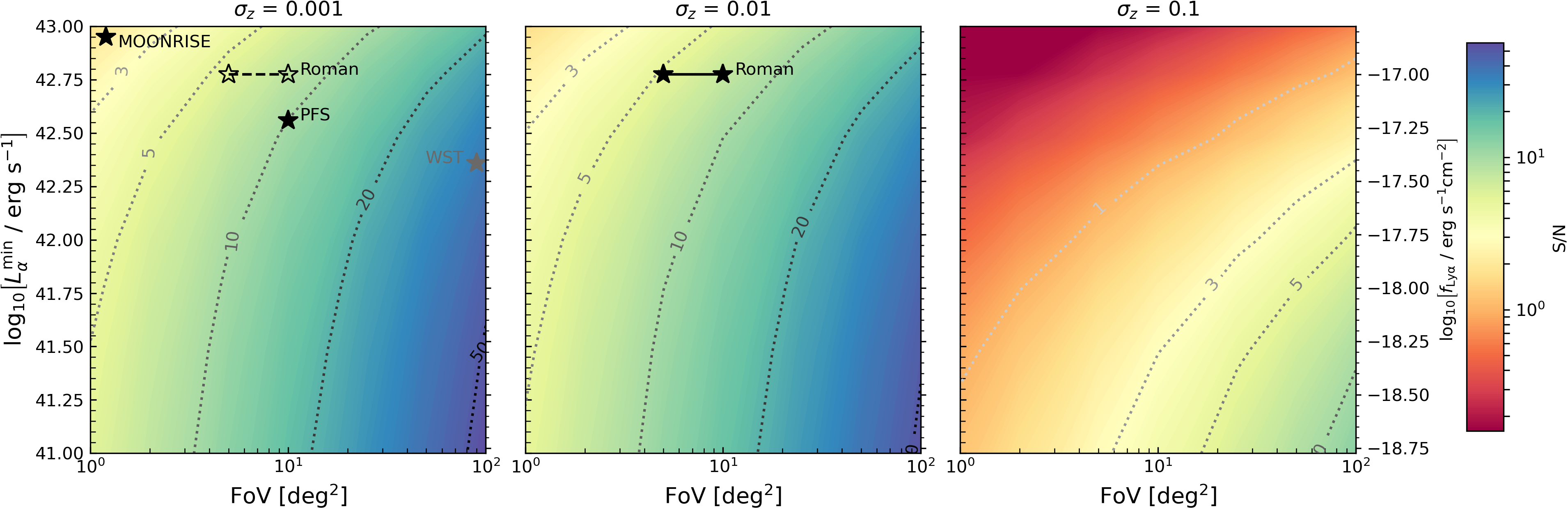}}
  \caption{Total S/N of the 21cm--LAE cross-power spectrum in the {\sc mhdec} reionisation scenario as a function of survey FoV and the minimum Lyman-$\alpha$ luminosity, $L_\alpha^\mathrm{min}$, shown for spectroscopic (left), grism (centre) and photometric-like (right) surveys at $z=7$. The 21cm signal noise is computed assuming the SKA-Low1 AA4 antenna layout, $\sim100~h$ integration time, and an optimistic foreground model. Black and grey stars mark potential LAE surveys at $z\simeq7$ that could be cross-correlated with SKA 21cm data (see also Tab.~\ref{tab_surveys}).}
     \label{fig_totalSNR_paramspace_MHDEC_opt_100h}
\end{figure*}

\begin{figure*}
\resizebox{\hsize}{!}
        {\includegraphics{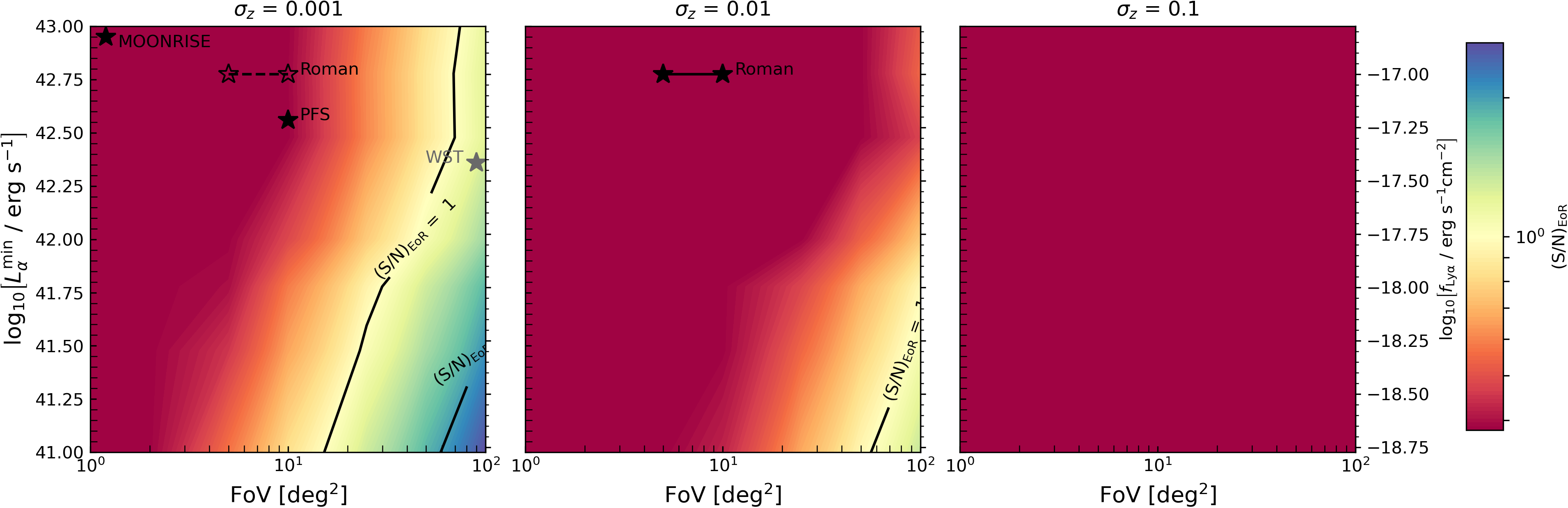}}
  \caption{Total S/N (black lines and coloured contours) for distinguishing the {\sc mhdec} and {\sc mhinc} reionisation scenarios using the $z=7$ 21cm--LAE cross-power spectra as a function of the survey FoV and the minimum Lyman-$\alpha$ luminosity, $L_\alpha^\mathrm{min}$, shown for spectroscopic (left), grism (centre) and photometric-like (right) surveys. The 21cm signal noise is computed assuming the SKA-Low1 AA4 antenna layout, $\sim100~h$ integration time, and a moderate foreground model. Black and grey stars mark potential LAE surveys at $z\simeq7$ that could be cross-correlated with SKA 21cm data (see also Tab.~\ref{tab_surveys}).}
     \label{fig_totalSNR_paramspace_reionisation_scenarios_mod_100h}
\end{figure*}

\begin{figure*}
\resizebox{\hsize}{!}
        {\includegraphics{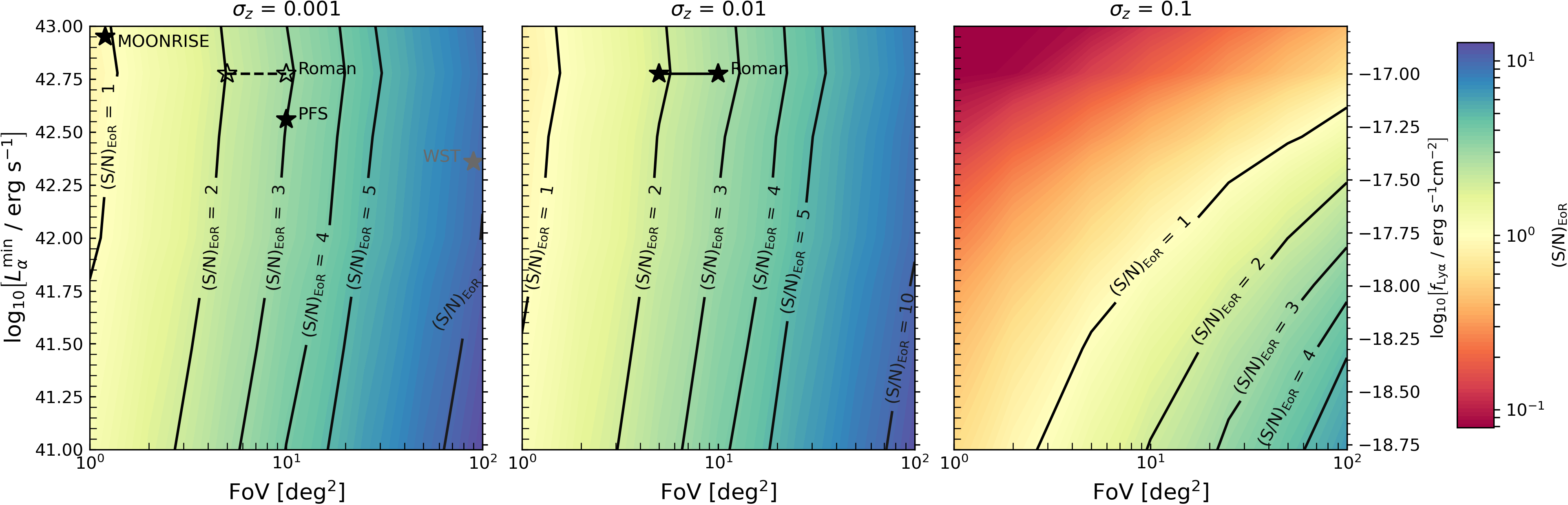}}
  \caption{Total S/N (black lines and coloured contours) for distinguishing the {\sc mhdec} and {\sc mhinc} reionisation scenarios using the $z=7$ 21cm--LAE cross-power spectra as a function of the survey FoV and the minimum Lyman-$\alpha$ luminosity, $L_\alpha^\mathrm{min}$, shown for spectroscopic (left), grism (centre) and photometric-like (right) surveys. The 21cm signal noise is computed assuming the SKA-Low1 AA4 antenna layout, $\sim100~h$ integration time, and an optimistic foreground model. Black and grey stars mark potential LAE surveys at $z\simeq7$ that could be cross-correlated with SKA 21cm data (see also Tab.~\ref{tab_surveys}).}
     \label{fig_totalSNR_paramspace_reionisation_scenarios_opt_100h}
\end{figure*}

We briefly assess how the S/N values change for detecting the 21cm–LAE cross-power spectrum at $z=7$, and the survey requirements for distinguishing the {\sc mhdec} and {\sc mhinc} reionisation scenarios when the SKA-Low1 integration time is reduced from 1000~h to 100~h. From Figs.~\ref{fig_totalSNR_paramspace_MHDEC_mod_100h} and \ref{fig_totalSNR_paramspace_MHDEC_opt_100h}, it is apparent that the S/N for detecting the cross-power spectrum drops by roughly $30$\% for both {\sc mhdec} and {\sc mhinc} scenarios. Similarly, the S/N values for distinguishing the two reionisation scenarios decrease by a comparable amount, as shown in Figs.~\ref{fig_totalSNR_paramspace_reionisation_scenarios_mod_100h} and \ref{fig_totalSNR_paramspace_reionisation_scenarios_opt_100h}. Although the integration time is reduced by an order of magnitude, the S/N drop is relatively modest. This weak dependence arises because the high signal-to-noise modes are the larger-scale modes (smaller $k$ values), where cosmic variance dominates the cross-power uncertainties.

\end{document}